\documentclass[article]{agujournal2019}
\usepackage{url} 
\usepackage{lineno}
\usepackage[inline]{trackchanges} 
\usepackage{soul}

\usepackage{timet,color}
\usepackage[labelfont=bf]{caption}
\usepackage{todonotes}
\usepackage{import}

\usepackage{bm}
\usepackage{bbm}
\usepackage{amsmath}
\usepackage{amssymb}
\usepackage{mathtools}
\allowdisplaybreaks

\usepackage[ruled,vlined]{algorithm2e}
\usepackage{scalefnt}
\usepackage{float}

\floatstyle{ruled}

\newfloat{algo}{thp}{lop}
\floatname{algo}{Algorithm}
\newcommand{\bsp}[1]{#1}
\draftfalse

\journalname{}

\begin{document}

\title{Bayesian Detectability of Induced Polarisation in Airborne Electromagnetic Data using Reversible Jump Sequential Monte Carlo}

%
%




\authors{L. Davies\affil{1}\affil{2}, A.Y. Ley-Cooper\affil{3}, M. Sutton\affil{1}\affil{2}, C. Drovandi\affil{1}\affil{2}}
\affiliation{1}{School of Mathematical Sciences, Queensland University of Technology, Brisbane QLD \emph{4000}, Australia}
\affiliation{2}{Centre for Data Science, Queensland University of Technology, Brisbane QLD \emph{4000}, Australia}
\affiliation{3}{Geoscience Australia, Canberra ACT \emph{2609}, Australia}








\begin{keypoints}
\item We introduce a reduced model space approach via decoupled layered model parameters for inference of  induced polarisation (IP) models versus conductive-only models fitted to airborne electromagnetic (AEM) data.
\item For model and parameter inference we develop an adaptive static sequential Monte Carlo algorithm with reversible jump Markov chain Monte Carlo proposals (RJSMC).
\item We successfully apply RJSMC to perform airborne induced polarisation (AIP) detectability on data from a large survey that demonstrates spatial continuity.
\end{keypoints}

%
%

%
%


\begin{abstract}
Detection of induced polarisation (IP) effects in airborne electromagnetic (AEM) measurements does not yet have an established methodology. This contribution develops a Bayesian approach to the IP-detectability problem using decoupled transdimensional layered models, and applies an approach novel to geophysics whereby transdimensional proposals are used within the embarrassingly parallelisable and robust static Sequential Monte Carlo (SMC) class of algorithms for the simultaneous inference of parameters and models. Henceforth referring to this algorithm as Reversible Jump Sequential Monte Carlo (RJSMC), the statistical methodological contributions to the algorithm account for adaptivity considerations for multiple models and proposal types, especially surrounding particle impoverishment in unlikely models. Methodological contributions to solid Earth geophysics include the decoupled model approach and proposal of a statistic that use posterior model odds for IP detectability. A case study is included investigating detectability of IP effects in AEM data at a broad scale.
\end{abstract}

\section*{Plain Language Summary}
Electromagnetic models for solid-Earth geophysics often make assumptions to ease computation. These assumptions may hold in the majority of cases, however in cases where it is impossible to explain certain problematic data using such models it is necessary to either revisit the model assumptions or to consider empirical model approximations. In this research, the problematic data contains significant anomalous measurements which are hypothesised to be due to the presence of a phenomenon known as induced polarisation (IP). This phenomenon is possibly explained using an empirical extension to current epistemological physical models given sufficient statistical evidence. It is the purpose of this research to introduce a rigorous statistical methodology for detecting when it is more likely that the empirical model will explain the data versus the epistemological model. This methodology is tested on artificial and real-world data, demonstrating the applicability and usefulness of the approach.

\section{Introduction}

Negative or steeply-decaying measurements of magnetic flux density in a concentric-loop airborne electromagnetic (AEM) system are usually inexplicable in electromagnetic models that ignore intrinsic chargeability. Despite wide acknowledgement of the possibility that these and other anomalies are caused by induced polarisation (IP) in subsurface materials \cite {kratzer_induced_2012}, confirmation of these hypotheses for airborne data is currently an open area of research. Moreover, the influence of such IP effects is not limited to producing negative measurements; significant distortions in off-time transients can present in any number of ways that can produce incorrect conductivity values in non-IP ground models \cite{viezzoli_airborne_2020}.

Deterministic methods for inverting time-domain AEM data with hypothesised IP effects generally do not consider parameter or model uncertainty, and as such do not directly provide a means for model inference. Approximate methods for model selection such as the Akaike Information Criterion (AIC) \cite{akaike_new_1974} are not robust to the pathological posteriors often found in layered-Earth geophysical models. To date, the most advanced Bayesian sampling approach applied to the Cole-Cole IP model \cite{cole_dispersion_1941} has been Markov Chain Monte Carlo (MCMC) for within-model parameter inference of non-airborne time and frequency domain data \cite{ghorbani_bayesian_2007}. Related work in conductive-only electromagnetic models applied to airborne data has advanced as far as transdimensional inference via various implementations of reversible jump Markov chain Monte Carlo (RJMCMC) \cite{brodie_transdimensional_2012,minsley_trans-dimensional_2011,hawkins_trans-dimensional_2017} and parallel-tempering RJMCMC \cite{blatter_trans-dimensional_2018}. It is a natural progression to consider Bayesian transdimensional inference methods for IP detection in airborne data in a similar manner.

Induced polarisation in AEM data, hitherto referred to as airborne induced polarisation (AIP), has attracted recent research interest posing the question of detecting such effects. This has been driven by the fact that ground-based IP methods comprise some of the most widely used techniques in mineral exploration, not only for the discovery of many anomalous mineralisation prospects due to their chargeable response \cite{meju_geoelectromagnetic_2002}, but also for the increasing ability to identify subsurface materials and mineralogy (\citeA{merriam_induced_2007}, \citeA{qi_induced_2018}, \citeA{feng_quantifying_2020}). Approaches using thresholds of re-parameterisations 
\cite{fiandaca_re-parameterisations_2018}, modelling of 3D IP effects \cite{nunes_detectability_2019}, and detection of IP using various approaches \cite{kang_detecting_2019,viezzoli_robust_2021} are some examples. However, Bayesian inference on the detectability of IP effects in AEM data has not been considered. This contribution will demonstrate how AIP detectability can be framed as a tractable model selection problem which can be accomplished using Bayesian methods. In addition, this work introduces the novel application of static Sequential Monte Carlo with an RJMCMC mutation kernel to parameter and model inference in AIP. First formally addressed in \citeA{del_moral_sequential_2006,jasra_interacting_2008} and subsequently identified in \citeA{zhou_toward_2016} as the SMC1 algorithm, this work will expand on the implementation of adaptive considerations in such an algorithm both generally and specifically for layered Earth models. It will henceforth be referred to as the Reversible Jump Sequential Monte Carlo (RJSMC) algorithm.

Application of SMC methods to geophysics problems is still relatively new. Recent work by \citeA{amaya_adaptive_2021} demonstrates how adaptive static SMC can be used to an advantage when needing to sample from complex priors. Earlier work by \citeA{dettmer_sequential_2011} demonstrates a completely different approach whereby in an application to sequential inference on geoacoustic survey lines the posterior distribution of a previous model in the sequence is input as the initial distribution for inference on parameters of the next model. Whilst such approaches are similar by name, the algorithms of \citeA{amaya_adaptive_2021} and \citeA{dettmer_sequential_2011} are configured to target different statistical quantities. Our article contributes to the growing literature demonstrating that RJSMC is a viable alternative to other popular methods such as RJMCMC and parallel tempering \cite{swendsen_replica_1986} in Geophysics. In addition, we present for the first time an implementation of transdimensional proposals designed for layered-Earth models in the SMC framework.

In comparison to other particle-population methods such as Population MCMC \cite{jasra_population-based_2007} and parallel tempering \cite{swendsen_replica_1986}, SMC algorithms are based on a methodology of sequentially sampling from a sequence of probability distributions on a common space \cite{del_moral_sequential_2006}. These probability distributions are approximated using a cloud of weighted random samples, or particles, where the process of moving to the next distribution is via a combination of importance sampling, resampling, and a specifically chosen mutation or propagation kernel. SMC has several benefits: the embarrassingly-parallel mutation of $N$ particles rather than reliance on one or few sequentially computed chains in other MCMC methods translates easily to parallel computing architectures; the algorithm is robust to high-dimensional multi-modal, often pathological posteriors (referred to in \citeA{ellis_inversion_1998} as non-uniqueness in AEM geophysics models); the availability of the particle approximation to the current target distribution allows for novel tuning possibilities; and a trivial additional output of SMC is the estimation of otherwise intractable normalising constants.

By using an MCMC kernel for particle mutation in combination with an artificial sequence of distributions (such as likelihood annealing) SMC can be readily applied to static parameter estimation problems such as AIP. While the use of SMC for model selection in static problems can take several forms as explored in \citeA{zhou_toward_2016}, this contribution will focus on an implementation that returns the joint posterior of parameters and models, facilitating a straightforward application of Bayesian model selection for the detection of AIP effects.

This paper is divided into three sections following this introduction. Section \ref{methodologysection}, titled {\em Methodology}, describes a modelling approach that employs two decoupled layered-Earth models to reduce the size of the model space, the application of RJSMC to the IP detectability problem in AEM data, where we introduce the Bayes Factor Induced Polarisation Detectability (BFIPD) statistic. Section \ref{sectioncompalgo} describes the computational algorithms for SMC and RJSMC, and proposes adaptive considerations for the latter. Section \ref{casestudiessection}, titled {\em Case Studies}, applies the described methodology and RJSMC algorithm to a synthetic study and subsequently to a real data set in Colorado USA demonstrating the spatial continuity of a novel IP detection statistic (described in Section \ref{bfdetection}).

\section{Methodology}\label{methodologysection}
\begin{figure}[!t]
    \includegraphics[width=1.0\textwidth]{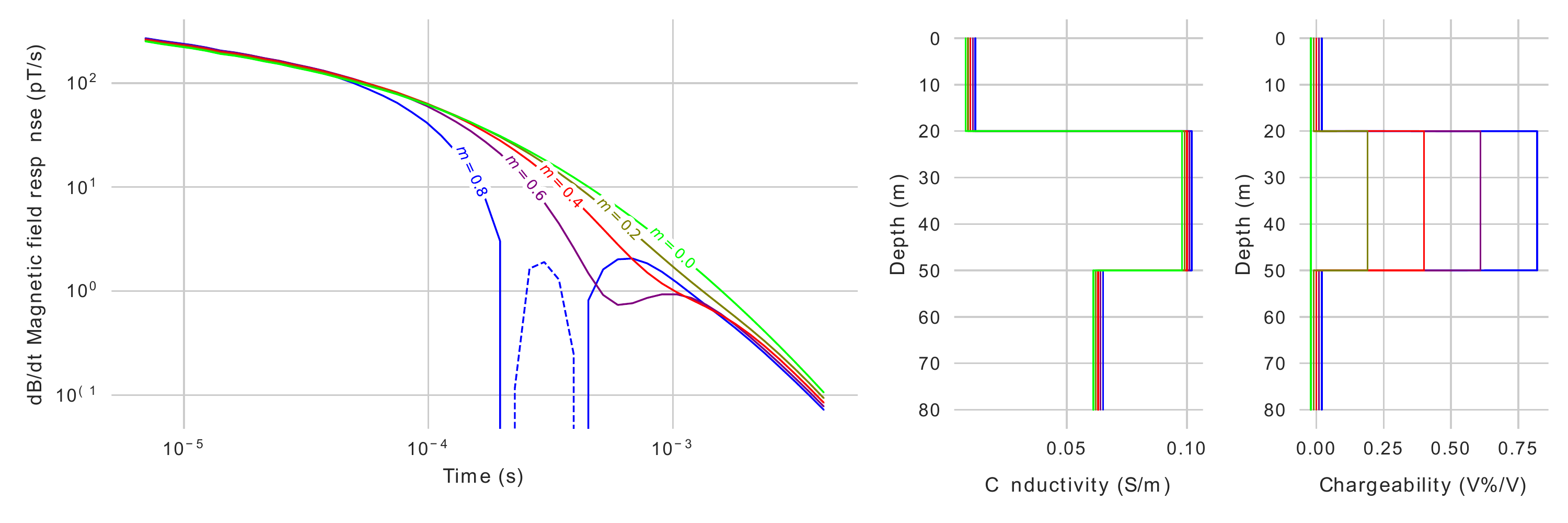}
    \caption[font=small]{Synthetic data produced using the Cole-Cole \cite{cole_dispersion_1941} parameterisation for an AEM layered-Earth model. The second layer intrinsic chargeability, $m$, varies from 0.0 to 0.8 to demonstrate the effect on the magnetic field response. Conductivity parameters are constant across models. Negative log values when $m \ge 0.8$ are plotted with a dashed line.}
    \label{figccsyn}
\end{figure}

For any flight location and geometry of an airborne electromagnetic (AEM) system, the response to the magnetic flux density response can be computed analytically using the 1D layered-Earth model approximation which employs a Hankel transform and propagation matrix method \cite{ward_4_1988}.  This approximation use the quasi-static assumption to reduce the Helmholtz wave equations to more tractable diffusion equations. In doing so, the dielectric permittivity term that models chargeability is removed. To re-introduce this term, the empirical Cole-Cole equation can be formulated (following \citeA{seigel_mathematical_1959}) using low and high frequency conductivity terms $\sigma_s$ and $\sigma_\infty$ respectively with intrinsic chargeability defined as $m=\frac{\sigma_\infty-\sigma_s}{\sigma_\infty}$. This gives the following form for complex conductivity in terms of the high-frequency conductivity $\sigma_\infty$, frequency $\omega$, time delay constant $\tau$ and frequency dependence $c$:
\begin{align*}
       \hat{\sigma}(\omega)=\sigma_\infty\bigg(1 - \frac{m}{1+(i\omega\tau)^{c}}\bigg).
\end{align*}
Although other formulations exist (such as that of \citeA{pelton_mineral_1978}), they are not considered here since such models can be computed from the above via a bijective transformation \cite{tarasov_use_2013}. This work considers only time domain elecromagnetic (TDEM) airborne data for inversion using independent 1D layered-Earth models for each AEM sounding. A depiction of models of varying chargeability and the resultant off-time TDEM synthetic responses computed using this mathematical model is in Figure \ref{figccsyn}.
Since the number of layers is not known \emph{a-priori}, a transdimensional approach is adopted to infer the number of layers. The Bayesian approach for this described in \citeA{malinverno_parsimonious_2002} introduced RJMCMC proposals for the birth and death of layer interfaces, each being comprised of a depth (or thickness) and one or more electromagnetic parameters. All parameters and associated prior distributions are identified in Section \ref{sectionpriors}, however in the next section they will all be referred to in concatenated vector form using the symbol $\bm{\theta}$.

\subsection{Decoupled Layered Models}
\begin{figure}
\includegraphics[width=1.0\textwidth]{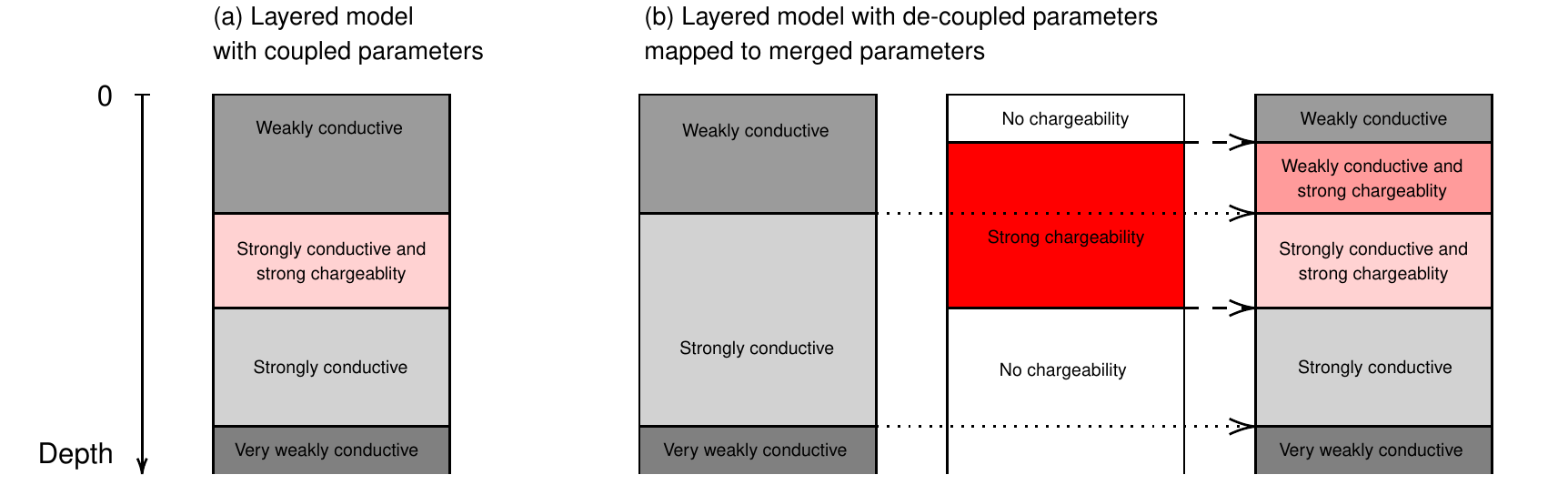}
\caption{(a) A single transdimensional parameter that models both conductivity and chargeability with respect to depth. Change to layer interface depths will change both conductivity and chargeability profiles because they are coupled. (b) Two de-coupled transdimensional parameters each separately modelling conductivity and chargeability with respect to depth. These are merged via a direct sum before computing the mathematical model.}
\label{fig:layeredmodeldecoupled}
\end{figure}

A primary concern for algorithmic complexity is the size of the model space. If we desire inference on the detectability of Cole-Cole chargeability in every conductive layer in a 1D model then we are implicitly using binary variables to model the inclusion of such parameters in each layer. We call this a {\em coupled} layered-model, 
where for $k$ layers this effectively poses a model space $\mathcal{M}$ of exponential cardinality.
Instead of investigating detectability of IP in this manner, we propose to {\em de-couple} the conductivity and Cole-Cole chargeability layer interfaces. This approach has been used in previous geophysical applications such as joint inversion of physically unrelated data in a single model \cite{piana_agostinetti_flexible_2018}. In this application it permits a more flexible and much smaller model space $\big|\mathcal{M}_{\mathrm{decoupled}}\big|=k_\kappa \times k_\lambda$ where $k_\kappa, k_\lambda$ signify the number of independent conductivity and Cole-Cole chargeability layers respectively. A comparison of a decoupled design with a coupled design is visualised in Figure \ref{fig:layeredmodeldecoupled}. The reduction of the model space cardinality is imperative when considering the inference approach we will take in the following sections whereby we begin by sampling the priors of each model.

The next subsection will introduce Bayesian inference of parameters and models, followed by a Bayesian description of the decoupled model for conductivity and chargeability. Subsequently, computational algorithms for inference on these parameters and model parameterisation will be discussed, including an introduction of the novel RJSMC sampler for which the decoupled model design is suited due to the reduced size of the model space.

\subsection{Bayesian inference of parameters and models}\label{sectionbayesianmodelselection}
Bayesian inference of parameters $\bm{\theta}$ given data $\bm{y}$, likelihood $\mathcal{L}(\bm{y}|\bm{\theta})$ (explained in Section \ref{sectionlikelihood}) and prior $p(\bm{\theta})$ (see Section \ref{sectionpriors}) is found via application of Bayes' theorem
\begin{align*}
    \pi(\bm{\theta}|\bm{y})
    &=
    \frac
    {\mathcal{L}(\bm{y}|\bm{\theta})p(\bm{\theta})}
    {\mathcal{Z}(\bm{y})},
\end{align*}
\noindent where the denominator term, known as the marginal likelihood or normalising constant, is the integral
\begin{align*}
    \mathcal{Z}(\bm{y})=\int_{\bm{\theta}}\mathcal{L}(\bm{y}|\bm{\theta})\pi(\bm{\theta})d\bm{\theta},
\end{align*}
\noindent which is typically computationally intractable in a  non-trivial number of dimensions. This term can be used for model selection using Bayes Factor \cite{kass_bayes_1995} (a quantity we will use in later sections) where between two contending models $k_1,k_2$ it is the ratio
\begin{align*}
    \mathcal{B}_{1,2}&=
    \frac
                {\int p(\bm{\theta}_{k_1}|k_1)\mathcal{L}(\bm{y}|\bm{\theta}_{k_1},k_1)d\bm{\theta}_{k_1}}
                {\int p(\bm{\theta}_{k_2}|k_2)\mathcal{L}(\bm{y}|\bm{\theta}_{k_2},k_2)d\bm{\theta}_{k_2}}
    =\frac
    {\mathcal{Z}(\bm{y}|k_1)}
    {\mathcal{Z}(\bm{y}|k_2)}.
\end{align*}
For inference over the joint space of models $k\in\mathcal{M}$ and parameters $\bm{\theta}_{k}$ we extend the above approach to express the posterior as
\begin{align*}
    \pi(\bm{\theta}_{k},k|\bm{y})
    &=
    \frac
    {\mathcal{L}(\bm{y}|\bm{\theta}_{k},k)p(\bm{\theta}_{k}|k)p(k)}
    {\sum_{k'\in \mathcal{M}}\mathcal{Z}(\bm{y}|k')},
\end{align*}
\noindent where the $\mathcal{Z}$ terms in the denominator will be henceforth defined as
\begin{align*}
    \mathcal{Z}_{k}=\mathcal{Z}(\bm{y}|k)&=\int_{\bm{\theta}_{k}}\mathcal{L}(\bm{y}|\bm{\theta}_{k},k)p(\bm{\theta}_{k}|k)p(k)d\bm{\theta}_{k}.
\end{align*}
If the above joint posterior of models and parameters is available, we can use Bayes' theorem to express Bayes Factor in terms of the ratios of posterior model marginal densities (also called the {\em posterior model odds}) and models priors
\begin{align*}
    \mathcal{B}_{1,2}&=\frac{\mathcal{Z}_{k_{1}}}{\mathcal{Z}_{k_{2}}}=\frac{\pi(k_1|\bm{y})}{\pi(k_2|\bm{y})}\frac{\pi(k_2)}{\pi(k_1)}.
\end{align*}

\subsection{Bayesian inference of induced polarisation models}
This subsection will employ the Bayesian approach to specifying IP and non-IP models and the inference of such models and parameters with respect to AEM data.

\subsubsection{Parameter Priors}\label{sectionpriors}

Using the decoupled layered model design, we identify the following parameter priors (using the indices $i,j$ to denote the $i^{th}$ conductive layer and $j^{th}$ chargeable layer respectively). Note that we use the notational convenience $\phi=\log_{10}\sigma$.
\begin{align*}
    \text{Number of conductive layers : } \kappa &\sim \mathcal{U}\{0,\dots,\kappa_{max}\},\\
    \text{Number of chargeable layers : } \lambda &\sim \mathcal{U}\{0,\dots,\lambda_{max}\},\\ 
    \text{Background Log10 conductivity : } \theta_{\phi,b} &\sim \mathcal{U}(-4,2),\\
    i^{th}\text{-layer Log10 conductivity : } \theta_{\phi,i} &\sim \mathcal{U}(-4,2),\\
    i^{th}\text{-layer Interface Depth : } \theta_{z_\sigma,i} &\sim \frac{\kappa!}{z_{max}^{\kappa}},\\
    \text{Background Instrinsic Chargeability : }\theta_{m,b}&=0,\\
    j^{th}\text{-layer Intrinsic Chargeability : } \theta_{m,j} &\sim \mathcal{U}(0,1),\\
    j^{th}\text{-layer Interface Depth : }  \theta_{z_m,j} &\sim \frac{\lambda!}{z_{max}^{\lambda}},\\
    \text{Time Constant : } \theta_{\tau} &\sim \mathcal{U}(0,1),\\
    \text{Frequency Dependence : } \theta_{c} &\sim \mathcal{U}(0,1).
\end{align*}
Using bold notation $\bm{\theta}_{\phi}$ to denote all parameters $\{\theta_{\phi,i}\}_{i=1}^{\kappa}$ (and similarly for $\bm{\theta}_{z_\sigma}$, $\bm{\theta}_{m}$, and $\bm{\theta}_{z_m}$) the full prior distribution becomes
\begin{align*}
    p(\kappa,\lambda,\theta_{\phi,b},\bm{\theta}_{\phi},\bm{\theta}_{z_\sigma},\bm{\theta}_{m},\bm{\theta}_{z_m},\theta_{\tau},\theta_{c})&=p(\kappa)p(\bm{\theta}_{\phi},\bm{\theta}_{z_\sigma}|\kappa)p(\lambda)p(\bm{\theta}_{m},\bm{\theta}_{z_m}|\lambda)p(\theta_{\phi,b})p(\theta_{\tau})p(\theta_{c}).
\end{align*}
The parameter vector in full, using the pair $\kappa,\lambda$ to identify the model, is
\begin{align*}
    \bm{\theta}_{\kappa,\lambda}&=\big\{\theta_{\phi,b},\theta_{\phi,1},\dots,\theta_{\phi,\kappa},\theta_{z_\sigma,1},\dots,\theta_{z_\sigma,\kappa},\theta_{m,1},\dots,\theta_{m,\lambda},\theta_{z_m,1},\dots,\theta_{z_m,\lambda},\theta_{\tau},\theta_{c}\big\}.
\end{align*}
Note that $\theta_\tau$ and $\theta_c$ Cole-Cole parameters are global for all chargeable layers. This could feasibly mean that there theoretically exist chargeable models that poorly fit the above parameterisation; however, in practical examples this has not yet been an issue.

\subsubsection{The Likelihood}\label{sectionlikelihood}

The data $\bm{y}$ is represented by an time-series array of response values, usually the time rate-of-change of the magnetic field ($\frac{dB}{dt}$) with units in picoteslas per second (pT/s), which for conductive only models are usually positive. Acquisition of data occurs during the off-time phase of the periodic waveform for the transmitted current. For example, the VTEM ET system \cite{eadie_vtem_2018} begins data acquisition from 5 microseconds after the start of the off-time period and continues sampling for up to an additional 15 milliseconds. Data is then downsampled to a time series of gates which are typically exponentially-spaced with respect to increasing time. Gates are indexed in the below using the square bracket notation $y[i]$. Visualisations of synthetic data were introduced at the beginning of this section on methodology using various example $\bm{\theta}$ parameterisations in Figure \ref{figccsyn}.

The mathematical {\em forward model} for an AEM system is a non-injective (and often non-surjective) map from the 1D layered-Earth model parameterisation of conductivity and chargeability parameters, $\bm{\theta}$, to the data space of off-time magnetic flux density responses, denoted $\bm{y}^{\bm{\theta}}$. The likelihood $\mathcal{L}(\bm{\theta}|\bm{y})$ is a multivariate Gaussian $\mathcal{N}(\bm{y};\bm{y}^{\bm{\theta}},\bm{\Sigma}_{\bm{y}^{\bm{\theta}}})$ computed in the synthetic data space. Typical methods for determining $\bm{\Sigma}_{\bm{y}^{\bm{\theta}}}$, i.e. the system noise, use an empirical approximation from high-altitude measurements where the effects of the ground are not present. This research employs the empirical model of \citeA{green_estimating_2003} where $\bm{\Sigma}_{\bm{y}^{\bm{\theta}}}$ is a diagonal covariance matrix where each diagonal element is the summation in quadrature of additive noise $\epsilon_{\mathrm{AN}}$ for that time window and multiplicative noise $\epsilon_{\mathrm{MN}}$, that is
\begin{align*}
    \bm{\Sigma}_{\bm{y}^{\bm{\theta}}}^{(i,j)}&=\begin{cases}
    (\epsilon_{\mathrm{AN}}[i])^2 + (\epsilon_{\mathrm{MN}}\times y^{\bm{\theta}}[i])^2, &i=j,\\
    0, &i\neq j.
    \end{cases}
\end{align*}

Such an approach requires high-altitude calibration lines to be flown immediately prior to data acquisition, and such calibration determines a per-time-window value for $\epsilon_{\mathrm{AN}}$ and an overall value for $\epsilon_{\mathrm{MN}}$, where typical values for $\epsilon_{\mathrm{MN}}$ are in the order of $5\%$. The empirical additive and multiplicative noise model is adequate for data where only conductivity parameters are of interest. However, in data with one or several significant zero-crossings in middle to late time windows, the additive noise will become the dominant influence in the likelihood. In this research, it was found that ``small" additive noise resulted in poor estimation of the parameter posteriors; this observation was consistent with other Bayesian research in transdimensional geophysics methods \cite{bodin_transdimensional_2012} which notes that the variance of data noise strongly affects the shape of the posterior. Since the AEM system noise models were not necessarily constructed with Bayesian sampling methods for IP in mind, the additive noise was kept above $10^{-3}$ pT/s to ensure that reasonable estimates of the posterior were feasible. It would be the subject of future research to determine a noise model that is parametric such as that in \citeA{bodin_transdimensional_2012}.

\subsubsection{A statistic for Bayesian detectability of induced polarisation}\label{bfdetection}
Inference of detectability of IP effects in this work is via Bayes factor using the expected probabilities of chargeable models and non-chargeable models. Denoting the chargeable versus non-chargeable estimate of Bayes factor as $\widehat{\mathcal{B}}_{\kappa\lambda}$, where $k=\kappa$ is a non-chargeable model and $k=\lambda$ is a chargeable model, we define the Bayes Factor Induced Polarisation Detectability (BFIPD) statistic to be the grouped Bayes Factor
\begin{align*}
    \widehat{\mathcal{B}}_{\kappa\lambda} &= \frac{|\mathcal{M}_\mathrm{\kappa}|\sum\limits_{\lambda\in\mathcal{M}_\mathrm{\lambda}}\widehat{\pi}(\lambda|\bm{y})/ p(\lambda)}{|\mathcal{M}_\mathrm{\lambda}|\sum\limits_{\kappa\in\mathcal{M}_\mathrm{\kappa}}\widehat{\pi}(\kappa|\bm{y})/ p(\kappa)}.
\end{align*}
Using $\log\widehat{\mathcal{B}}_{\kappa\lambda}$, a chargeable model is more likely when this value is greater than zero. Applied to real data, the third section in Figure \ref{figconcharprofile} demonstrates how this quantity can be used to detect chargeability in Earth materials assuming no other anomalous effects.

\section{Computational algorithms for Bayesian model inference}
\label{sectioncompalgo}
The canonical approach for inference of $\pi(\bm{\theta}_k,k|\bm{y})$ is to implement a RJMCMC algorithm that proposes on the space of $k$ as well as $\bm{\theta}$. Whilst such algorithms asymptotically converge to the posterior, there are several concerns. One is the difficulty of designing a well-mixing RJMCMC proposal, another is the lack of parallelisability due to the dependent nature of the algorithm. To solve the first, either the various adaptive proposals that conform to rules of diminishing adaptation \cite{haario_adaptive_2001} or proposals that use local derivative information (\citeA{roberts_langevin_2002}, \citeA{girolami_riemann_2011}) have shown success for within model MCMC, but these do not always translate well for RJMCMC proposal design. For the second concern, frameworks such as parallel tempering \cite{swendsen_replica_1986} can be parallelised to a certain degree, but do not upscale well to take advantage of increasingly common very wide computing architectures.

The SMC family of algorithms can be adapted for static parameter inference (\citeA{chopin_sequential_2002}, \citeA{del_moral_sequential_2006}), with the added benefit of providing an estimate of the marginal likelihood \cite{del_moral_sequential_2006}. For parameter estimation in a single non-linear model, static SMC has been demonstrated to have advantages when posteriors are pathological. To date, applications of static SMC in solid Earth geophysics are currently sparse, however recent work by \citeA{amaya_adaptive_2021} demonstrates static SMC for within model parameter inference as well as the use of estimates of the marginal likelihood for model selection. In the following section, a static SMC algorithm will be introduced and followed by an implementation which uses transdimensional proposals as well as within-model proposals, which we call Reversible Jump Sequential Monte Carlo (RJSMC).

\subsection{Static Sequential Monte Carlo}\label{smcappendix}
In this section we will briefly state the algorithm in the common configuration where an MCMC kernel is used for mutation (the reader may be interested in the work of \citeA{dai_invitation_2020} for a more in-depth and up-to-date review of SMC samplers). Such a configuration ensures that the computation of incremental particle weights can be evaluated with linear time complexity without the need to compute or approximate an expensive integral \cite{del_moral_sequential_2006}. The first implementation of this approach was called Iterated Batch Importance Sampling \cite{chopin_sequential_2002}, which used a {\em data annealing} schedule for successive target distributions. For brevity, the below description does not include commonly employed adaptive schemes such as those employed in \citeA{fearnhead_adaptive_2013}, however such approaches are addressed in the following section for RJSMC.

Following from Bayes Theorem in section \ref{sectionbayesianmodelselection}, the posterior of parameters $\bm{\theta}$ is proportional to the likelihood $\mathcal{L}$ and prior $p(\cdot)$, that is
\begin{align*}
    \pi(\bm{\theta}|\bm{y})\propto \mathcal{L}(\bm{y}|\bm{\theta}) p(\bm{\theta}).
\end{align*}
In a static SMC algorithm, a sequence of distributions $\pi_t$, $t=0,\dots,T$ is specified that ``smoothly" transitions from a starting distribution, most commonly the prior $ p(\cdot)$, to the target posterior distribution $\pi(\cdot|\bm{y})$. For inference using sparse data (such as AEM time-domain data) a common choice for this sequence is {\em likelihood annealing}, where a  monotonic sequence $\{\gamma_t\}_{t=0}^{t=T}$ with $\gamma_0=0$ ascending to $\gamma_T=1$ defines the sequence of target distributions
\begin{align*}
    \pi_t(\bm{\theta}|\bm{y})=\frac{ \mathcal{L}(\bm{y}|\bm{\theta})^{\gamma_t} p(\bm{\theta})}{\mathcal{Z}_t}.
\end{align*}
The proportionality constant $\mathcal{Z}_t$ usually cannot be evaluated, so we instead define a tractable term $\eta_t$ as
\begin{align*}
\eta_{t}(\bm{\theta})=\mathcal{Z}_{t}\pi_t(\bm{\theta}|\bm{y})=\mathcal{L}(\bm{y}|\bm{\theta})^{\gamma_t} p(\bm{\theta}).
\end{align*}
The target distributions $\pi_t$ are approximately represented by a set of $N$ particles. During initialisation, the particles are drawn from the prior $\pi_0=p(\cdot)$. Following this, for each temperature $\gamma_t$, a three-step procedure of {\em importance-sampling}, {\em resampling}, and {\em mutation} is evaluated. A typical mutation kernel choice for static parameter inference is a target-invariant MCMC kernel, resulting in importance weights taking the incremental form 
\begin{align*}
            w_{t}^{(i)}&=W_{t-1}^{(i)}\frac{\eta_{t}(\bm{\theta}^{(i)})}{\eta_{t-1}(\bm{\theta}^{(i)})}=W_{t-1}^{(i)}\mathcal{L}(\bm{y}|\bm{\theta})^{\gamma_t-\gamma_{t-1}},
\end{align*}
\noindent where $\bm{\theta}^{(i)}$ is the parameter vector for the $i^{\mathrm{th}}$ particle, and the term $W_{t-1}^{(i)}$ represent normalised weights from the previous target, computed via
\begin{align*}
    W_t^{(i)}&=\frac{w_t^{(i)}}{\sum_{j=1}^N w_t^{(j)}},
\end{align*}
\noindent where $W_0^{(i)}=\frac{1}{N}$ for all $i$ particles. The resampling step simply draws $N$ new particles $\bm{\theta}^{(i)*}$ from the weighted representation of particles $\{W^{(i)}_t,\bm{\theta}^{(i)}\}_{i=1}^{N}$ and resets the normalised weights to $W_t^{(i)}=\frac{1}{N}$ for all $i$ particles.

As stated earlier, the mutation step uses a target-invariant MCMC kernel such as several iterations of random-walk Metropolis-Hastings \cite{hastings_monte_1970} to perturb each particle. The resulting particle will then be used as input for the next target $\pi_{t+1}$. Denoting particles for targets at steps $t$ and $t+1$ as $\bm{\theta}_{t}^{(i)}$ and $\bm{\theta}_{t+1}^{(i)}$ respectively, we write the mutation step as
\begin{align*}
        \bm{\theta}_{t+1}^{(i)}\sim K(\cdot|\bm{\theta}_{t}^{(i)*}).    
\end{align*}
This procedure is summarised in Algorithm \ref{ssmcalgo} in \ref{staticsmcalgorithm}. A by-product of this algorithm is the marginal likelihood estimate as found by the following trivial computation:
\begin{align*}
    \widehat{\mathcal{Z}}&=\prod_{t=1}^T\sum_{i=1}^{N} w^{(i)}_t .
\end{align*}
Some of the concerns mentioned at the end of Section \ref{sectionbayesianmodelselection} are addressed by using an SMC algorithm. By virtue of the particle approximation of $\pi_t$, proposals can be designed that take advantage of this information without violating any rules of diminishing adaptation. Secondly, the computation of $\mathcal{L}(\bm{y}|{\bm{\theta}})$ for $N$ particles can be performed independently in parallel, thus lending itself well to a distributed computing architecture. 

Another advantageous difference between the static SMC framework and traditional Markov Chain sampling methods is the implicit stopping condition that an adaptive SMC algorithm provides. Rather than determining the number of MCMC iterations {\em a-prior}, or relying on convergence criteria, SMC terminates after traversing the sequence of target distributions.

Whilst static SMC does produce an unbiased estimate of the normalising constant, it does require running an instance of the algorithm for each model. This general approach was recommended by \citeA{zhou_toward_2016}, but it ignores efficiencies that can be leveraged from existing research in Bayesian model selection for particular problems. In the case of layered-Earth models, the RJMCMC proposals first introduced by \citeA{malinverno_parsimonious_2002} and subsequently developed in \citeA{dosso_efficient_2014} are (assuming non-pathological likelihood conditions) capable of efficiently traversing a medium-sized model space and sampling high likelihood models more often than low likelihood models, thereby implicitly introducing a sampling efficiency that in practice yields posterior model probabilities with low variability. We intend to leverage this efficiency in the implementation of static SMC with RJMCMC proposals discussed in the next section, and apply this approach to a model space with cardinality approaching $10^2$. Such a broad prior of models is commonplace in exploration geophysics problems where there is usually a wide variation of possible posteriors with very little informative prior knowledge.

\subsection{Reversible Jump Sequential Monte Carlo}
The target density we wish to consider is the joint posterior of models and parameters
\begin{align*}
    \pi(\bm{\theta}_k,k|\bm{y})&\propto\mathcal{L}(\bm{y}|\bm{\theta}_k,k)p(\bm{\theta}_k|k)p(k).
\end{align*}
Based on the SMC1 archetype identified in \citeA{zhou_toward_2016}, we specify a static SMC algorithm using a likelihood-annealed sequence of target distributions where the annealing exponent sequence $\gamma_t$ is monotonically increasing on $[0,1]$. The below sequence of target distributions forms the basis of the RJSMC algorithm
\begin{align*}
    \pi_t(\bm{\theta}_k,k|\bm{y})&\propto\mathcal{L}(\bm{y}|\bm{\theta}_k,k)^{\gamma_t} p(\bm{\theta}_k|k) p(k),\   t=0,\dots,T.
\end{align*}
 If we denote the normalising constant defined in Section \ref{sectionbayesianmodelselection} for target density $\pi_t$ as $\mathcal{Z}_{t,k}$, we note that the above is proportional to $\sum_{k\in\mathcal{M}}\mathcal{Z}_{t,k}$. For convenience of notation, we introduce $k$ terms $\eta_{t,k}$ for each unnormalised target density
\begin{align*}
\eta_{t,k}(\bm{\theta}_{k})=\mathcal{Z}_{t,k}\pi_t(\bm{\theta}_k,k|\bm{y})=\mathcal{L}(\bm{y}|\bm{\theta}_k,k)^{\gamma_t} p(\bm{\theta}_k|k) p(k).
\end{align*}
Using this, and assuming an RJMCMC kernel is used for particle dynamics, we again choose to approximate the posteror of models and parameters $\pi_t$ with $N$ particles, and as such we define the importance weights for each particle, indexed by $i$ and for convenience the model $k$, as
\begin{align*}
            w_{t,k}^{(i)}&=W_{t-1,k}^{(i)}\frac{\eta_{t,k}(\bm{\theta}_{k}^{(i)})}{\eta_{k}(\bm{\theta}_{k}^{(i)})}=W_{t-1,k}^{(i)}\mathcal{L}(\bm{y}|\bm{\theta}_k)^{\gamma_t-\gamma_{t-1}},
\end{align*}
\noindent where we denote the number of particles representing $\pi_{t,k}(\bm{\theta}|k)$ by $N_{t,k}$. Note that since the algorithm is using particle approximations to such conditional densities, the total number of particles $N=\sum_{k\in\mathcal{M}}N_{t,k}$ should be set such that $N\ll|\mathcal{M}|$. The weights are then normalised such that they approximately represent the conditional density $\pi_t(\bm{\theta}_{t,k}|k,\bm{y})$. These normalised weights are given by
\begin{align*}
    W_{t,k}^{(i)}&=\frac{w_{t,k}^{(i)}}{\sum_{j=1}^{N_{t,k}}w_{t,k}^{(j)}}.
\end{align*}
Following the notation from the previous section on single-model SMC, parameters for the $i^{th}$ particle associated with target $k$ at steps $t$ and $t+1$ will henceforth be denoted $\bm{\theta}_{t,k}^{(i)}$ and $\bm{\theta}_{t+1,k}^{(i)}$ respectively. At initialisation, after sampling from the joint prior of model and parameters, the normalised weights are uniform, i.e. $W_{t,k}^{(i)}=1/N_{t,k}$ for all particles $i=1,\dots,N_{t,k}$ in model $k$. The same uniform initialisation is applied after the mutation step on the (now different) set of particles that represent the subsequent conditional target density $\pi_{t+1}(\bm{\theta}_{t+1,k}|k,\bm{y})$.

Since we are using an RJMCMC kernel, 
$N_{t,k}$ will change at the mutation step and as such it is not fixed for all $t$ in $\pi_t(\bm{\theta}_{t,k}|k,\bm{y})$.
It is a design choice of this algorithm to keep $N_{t,k}$ constant in the resample step, and a natural way to achieve this is to sample $\bm{\theta}_{t,k}^{(i)*}\sim\widehat{\pi}_{t}(\cdot|k)$ where $\widehat{\pi}_{t}(\cdot|k)$ is represented by the weighted particles $\{W^{(i)}_{t,k},\bm{\theta}_{t,k}^{(i)}\}_{i=1}^{N_{t,k}}$ from model $k$. We refer to this constraint as {\em within-model re-sampling}, and by observing this constraint a typical scheme such as multinomial or systematic resampling can be used. The complete adaptive algorithm is summarised in Algorithm \ref{rjsmcalgo}.

\begin{algorithm}[!h]
\caption{Adaptive Reversible Jump Sequential Monte Carlo}
\label{rjsmcalgo}
\SetAlgoLined
\DontPrintSemicolon
    \KwIn{TESS threshold proportion $\alpha_{\mathrm{TESS}}$, $c$ probability of mutation}
    \KwOut{$\{\bm{\theta}_{T,k}^{(i)}\}_{i=1}^{N_{t,k}}$ and $\widehat{\mathcal{Z}}_{T,k}$ for $k\in\mathcal{M}$}
    Initialise particles from the prior for $i=1,\dots,N$,$k\sim\pi_0(k)=p(k)$, $\bm{\theta}_{0,k}^{(i)}\sim\pi_0(\cdot|k)=p(\cdot|k)$.\\
    Initialise the particle weights for $i=1,\dots,N_{t,k}$, $k\in\mathcal{M}$, $W_{0,k}^{(i)}\leftarrow{N_{t,k}}^{-1}$.\\
    Set $t\leftarrow 0$, $\gamma_0\leftarrow 0$, and $\widehat{\mathcal{Z}}_{0,k}=1$ for $k\in\mathcal{M}$.\\
    \While{$\gamma_t \neq 1$} {
        Set $t\leftarrow t+1$\;
        Compute $\gamma_t$ using a bisection method such that $\widehat{\mathrm{TESS}}_{t}\approx \alpha_{\mathrm{TESS}}N$\;
        \textbf{\textit{Re-weight}} particles with normalised weights for $i=1,\dots,N_{t,k}$, $k\in \mathcal{M}$\begin{align*}
            w_{t,k}^{(i)}&\leftarrow W_{t-1,k}^{(i)}\mathcal{L}(\bm{y}|\bm{\theta}_{t,k}^{(i)})^{\gamma_t-\gamma_{t-1}}\\W_{t,k}^{(i)}&\leftarrow \frac{w_{t,k}^{(i)}}{\sum_{i=1}^{N_{t,k}} w_{t,k}^{i)}}\\\widehat{\mathcal{Z}}_{t,k}&\leftarrow \widehat{\mathcal{Z}}_{t-1,k}\sum_{i=1}^{N_{t,k}}w_{t,k}^{(i)}\end{align*}\;
        \textbf{\textit{Re-sample}} $N_{t,k}$ particles $\bm{\theta}_{t,k}^{(i)*}\sim\widehat{\pi}_{t}(\cdot|k)$ where $\widehat{\pi}_{t}(\cdot|k)$ is represented by the weighted particles $\{W^{(i)}_{t,k},\bm{\theta}_{t,k}^{(i)}\}_{i=1}^{N_{t,k}}$ for $k\in\mathcal{M}$\;
        \textbf{\textit{Adapt}} MCMC and RJMCMC proposals $q_{k\rightarrow k'}$ using weighted power posterior $\{\{\nu_{s,k}^{(i)},\bm{\theta}_{s,k}^{(i)}\}_{i=1}^{N_{t,k}}\}_{s=0}^{t}$  where\begin{align*}\nu_{t,s,k}^{(i)}&=\frac{\mathcal{L}(\bm{y}|\bm{\theta}_{s,k}^{(i)})^{\gamma_t} p(\bm{\theta}_{s,k}^{(i)})}{\frac{1}{t}\sum_{l=0}^t \mathcal{L}(\bm{y}|\bm{\theta}_{s,k}^{(i)})^{\gamma_l} p(\bm{\theta}_{s,k}^{(i)})(\widehat{\mathcal{Z}}_{l,k})^{-1}}\text{\ \ \ for\ \ \ } k\in\mathcal{M}\end{align*}\;
        Compute $R_t = \left \lceil{\frac{\log{c}}{\log{(1-p_{\mathrm{acc}}^{\mathrm{min}})}}}\right \rceil$ using trial mutations\;
        \textbf{\textit{Mutate}} particles $R_t$ times using an RJMCMC kernel with the $n_q$ adapted proposals\begin{align*}k'&\sim q_\mathcal{M}(\cdot|k),\\\bm{\theta}_{t+1,k'}^{(i)}&\sim q_{k\rightarrow k'}(\cdot|\bm{\theta}_{t,k}^{(i)*},k')
        \end{align*}\;
    }
\end{algorithm}

\subsubsection{Adaptive Considerations}
The configuration of a static SMC sampler over the joint posterior of models and parameters is not well-researched in terms of the sequence of target densities for more than one model and the particle counts between models. In this subsection we will discuss the adaptation of the sequence of targets $\pi_t$, and in subsequent subsections we will address the determination of the number of MCMC mutation steps, and overcoming particle impoverishment for the later target densities of unlikely models. 

In the single-model static SMC formulation of \citeA{schafer_sequential_2011}, next target density $\pi_{t+1}$ is specified adaptively using an estimate of the Effective Sample Size (ESS) \cite{kish_survey_1965}, where a predetermined threshold $\alpha$ is used to find $\gamma_{t+1}$ such that the resampling phase is initiated when the ESS of $\pi_{t+1}$ falls below $\alpha N$. Such a scheme can be implemented na\"ively by choosing a small step size $\delta$ (where $\gamma_{t+1}=\gamma_{t}+\delta$) and simply incrementing $t$ until the threshold is reached. Alternatively, a bisection method can be used to determine the next $\gamma_{t+1}$ \citeA{jasra_inference_2011}.

In the presence of multiple models, the ESS estimate for the conditioned density $\pi_t(\cdot|k)$ is
\begin{align*}
    \widehat{\mathrm{ESS}}_{t,k} &= \frac{1}{\sum_{i=1}^{N_{t,k}} \big(W^{(i)}_{t,k}\big)^2}.
\end{align*}
If there are $K$ models where $K>1$, a new condition is required for when to stop and resample/mutate since there are now $K$ ESS estimates. Given a threshold $0<\alpha<1$, such a condition can be met by taking a statistic of $\widehat{\mathrm{ESS}}_{t,k}$. A na\"ive approach would be to take the threshold condition $\min_k (\widehat{\mathrm{ESS}}_{t,k}/N_{t,k}) < \alpha$, but this exposes the algorithm to high variability of $\widehat{\mathrm{ESS}}_{t,k}$ in unlikely models. For this reason, we choose to have the threshold condition be dominated by the more likely models. First, we need to define the {\em normalised } effective sample size (NESS) for the conditional target density $\pi_t(\bm{\theta}|\bm{y},k)$ as
\begin{align*}
    \mathrm{NESS}_{t,k}&=\frac{1}{N_{t,k}}\mathrm{ESS}_{t,k}.
\end{align*}
Using this quantity, we choose our threshold condition to use what we define as the Total Effective Sample Size (TESS), which expressed in terms of $\mathrm{NESS}_{t,k}$ is
\begin{align*}
    \mathbb{E}[\mathrm{TESS}_t]&\vcentcolon=N\cdot\mathbb{E}[\mathrm{NESS}_{t,k}].
\end{align*}
It is shown in \ref{tessappendix} that an estimate of the TESS is simply
\begin{align*}
    \widehat{\mathrm{TESS}}_t&=\sum_{k\in\mathcal{M}} \widehat{\mathrm{ESS}}_{t,k}.
\end{align*}
The algorithm will resample when $\widehat{\mathrm{TESS}}_t<\alpha N$. Intuitively, one can reason that in low-likelihood models the variability of $\mathrm{ESS}_{t,k}$ would increase, but since such models are represented by proportionally fewer particles this variability does not translate to increased variability in the overall TESS. As such, this approach is in practice robust to low particle counts in the presence of unlikely models.

An estimate of the ratio of normalising constants between successive target distributions in an SMC algorithm with an MCMC kernel is given by the sum of the weights. When considering the same quantity for an RJSMC algorithm, we show in \ref{adaptiverjmcmcappendix} that the normalising constant estimate $\widehat{\mathcal{Z}}_{T,k}$ reduces to the same form considering particle weights for model $k$. This gives us 
\begin{align*}
    \widehat{\mathcal{Z}}_{T,k}=\prod_{t=0}^T\widehat{\frac{\mathcal{Z}_{t,k}}{\mathcal{Z}_{t-1,k}}}=\prod_{t=0}^T\sum_{i=1}^{N_{t,k}} w_{t,k}^{(i)},
\end{align*}
\noindent noting that $\mathcal{Z}_{0,k}=1$. From here it is possible to compute estimates of the ratios of normalising constants between models using
\begin{align*}
    \widehat{\mathcal{B}}_{1,2}&=
    \frac
    {\widehat{\mathcal{Z}}_{T,k_1}}
    {\widehat{\mathcal{Z}}_{T,k_2}}.
\end{align*}
Since we are using an RJMCMC kernel, another approach (which is adopted in this work) to estimate ratios of normalising constants is to take ratios of the empirical posterior model marginal density weighted by prior densities
\begin{align*}
    \widehat{\mathcal{B}}_{1,2}&=\frac{\widehat{\pi}_{T}(k_1|\bm{y})}{\widehat{\pi}_{T}(k_2|\bm{y})}\frac{ p(k_2)}{ p(k_1)},
\end{align*}
\noindent where
\begin{align*}
    \widehat{\pi}_T(k|\bm{y})&=\frac{1}{N}\sum_{i=1}^{N}\mathbb{I}(k^{(i)}=k)=\frac{N_{T,k}}{N},
\end{align*}
\noindent noting $k^{(i)}$ is the model indicator for particle $i$ at step $T$. Other precise quantities relating to ratios of normalising constants derived from RJMCMC posteriors are explored in \citeA{bartolucci_efficient_2006}. The motivating reason for adopting these latter approaches is to, if possible, reduce the posterior model-odds variability. This places the onus of reducing such variability on the RJMCMC proposal performance rather than the performance of the SMC sampler as a whole, which is a topic that would require further research as it is not central to this article. The form of the MCMC and RJMCMC proposals for the AIP inference application are discussed in the next subsection.

\subsubsection{Transdimensional and within-model adaptive proposals}\label{adaptiveproposals}
For this work, within-model MCMC proposals are an adaptive component-wise random-walk Gaussian proposal distribution which utilises the availability of particles approximating $\pi_{t,k}$ to adapt the search direction and step size. 

Typical component-wise proposals treat the parameter vector as independent components and randomly choose a component and then sample a value for that component. This approach relies on well-conditioned posterior for good performance. In a layered Earth model of electrical properties, the parameter space generally yields a pathological posterior, hence a na\"ive approach is inefficient. A simple approximate fit of a multivariate Gaussian $\mathcal{N}(\bm{\mu}_t,\bm{\Sigma}_t)$ to the particles $\bm{\theta}^{(i)}$, $i=1,\dots,N$ approximating $\pi_t$ opens the ability to use principle components as search directions. Taking the eigenvalue decomposition $\bm{\Sigma}_t=\bm{U}\bm{\Lambda}\bm{U}^T$ where $\bm{\Lambda}$ is diagonal, our component-wise proposal selects a random component/column $\bm{U}_r$ for the search direction and scales the step size by the corresponding eigenvalue $\bm{\Lambda}_r$.  A bisection algorithm on trial proposals is then used to further scale proposals to target a desired acceptance rate. In this work we used $0.44$ since it is accepted as the optimal acceptance rate for component-wise sampling of a standard multivariate Gaussian target \cite{roberts_optimal_2001}.

The cross-dimensional proposals use the RJMCMC framework, first proposed by \citeA{green_reversible_1995} and developed for the 1D layered-Earth model by \citeA{malinverno_parsimonious_2002}. The form of the proposals used in this work extend the above using a simple adaptive design constructed in \ref{adaptiverjmcmcappendix}. Two separate proposals are used, each mapping auxiliary variables to Earth property parameters $\theta_{\phi,i}$ and $\theta_{m,j}$ respectively.

\subsubsection{Adapting the number of MCMC steps for multiple proposals}

There is no set procedure for determining the number of MCMC iterations for effective mutation. Taking the approach introduced by \citeA{drovandi_estimation_2011}, the number of mutation steps can be determined simply as a function of the acceptance rate $R(p_{\mathrm{acc}})$ for a single proposal and a tuning parameter $c$. However, for RJSMC there are multiple proposal types as a result of using a birth/death RJMCMC pair of proposals and a within-model MCMC proposal. Fortunately, by simple extension of the approach taken in \citeA{drovandi_estimation_2011}, it is shown in \ref{multipleproposalR} that for $n_q$ proposals, the minimum acceptance rate
\begin{align*}
    p_{\mathrm{acc}}^{\mathrm{min}}=\min_{j\in n_q} p_{\mathrm{acc}}^{(j)}
\end{align*}\noindent will determine the minimum number of mutations required to ensure that all particles mutate with probability $1-c$. This results in the formula for determining the minimum number of mutation steps to be
\begin{align*}
R &= \left \lceil{\frac{\log{c}}{\log{(1-p_{\mathrm{acc}}^{\mathrm{min}})}}}\right \rceil.
\end{align*}
Other more recent work by \citeA{bon_accelerating_2021} adopts a generalised approach which adaptively chooses a proposal step size based on a target expected squared jumping distance (ESJD). Such an approach could be examined in the context of RJSMC samplers in future research.

\subsubsection{Overcoming particle impoverishment in unlikely models}
Cross-dimensional proposals that target $\pi_t(\bm{\theta}_k,k|\bm{y})$ will favour models with higher probability. Therefore, in a particle approximation of $\pi_t$ where $N$ is fixed, examples will occur where unlikely models are represented by few or no particles. In the case where na\"ive (RJ)MCMC proposals (i.e. those that do not adapt to $\pi_t$) are used, this is not necessarily an issue. However, in order to take advantage of the availability of $\pi_t$ for proposal tuning (see section \ref{adaptiveproposals} and \ref{adaptiverjmcmcappendix} for examples in this application), particle impoverishment becomes an important issue.

One approach to alleviate particle impoverishment is to re-use particles from previous target densities. Since we are considering a likelihood-annealed sequence of distributions, it is natural to consider the deterministic mixture recycling approach employed by \citeA{nguyen_efficient_2015} and further implemented in SMC by \citeA{south_sequential_2019}. 

The particle mixture weights representing the power-posterior at step $t$ for particles $i=1,\dots,N_{t,k}$, steps $s=0,\dots,t$,  and models $k\in\mathcal{M}$ are
\begin{align*}\nu_{t,s,k}^{(i)}&=\frac{\mathcal{L}(\bm{y}|\bm{\theta}_{s,k}^{(i)})^{\gamma_t} p(\bm{\theta}_{s,k}^{(i)})}{\frac{1}{t}\sum_{l=0}^t \mathcal{L}(\bm{y}|\bm{\theta}_{s,k}^{(i)})^{\gamma_l} p(\bm{\theta}_{s,k}^{(i)})(\widehat{\mathcal{Z}}_{l,k})^{-1}}\end{align*}
Since the normalising constant $\mathcal{Z}_{t,k}$ for the power posterior $\pi_t$ is not available, we use the RJSMC estimate $\widehat{\mathcal{Z}}_{t,k}$.

It is important to note that the above form for recycled particle weights is conditional on the model $k$, meaning that these weighted particles are suited only to fitting such conditional proposal densities and are not valid for representing the joint posterior of parameters and models. 

Whilst particle recycling has been used offline for estimation of posterior statistics, an online implementation for estimation of intermediate densities for proposal tuning is not known to the authors, and is thus proposed here as a novel step to mitigate particle impoverishment during the course of the algorithm.

\section{Case Studies}\label{casestudiessection}

This section will demonstrate the application of parameter and model inference via the RJSMC algorithm as applied to various synthetic and real-data examples. It will begin with a comprehensive synthetic case study demonstrating the performance of Bayes Factor detectability of chargeability, followed by inference for IP detectability in a 2D ground section of AEM line data.
\subsection{Synthetic Studies}

\begin{figure}
    \includegraphics[width=1.0\textwidth]{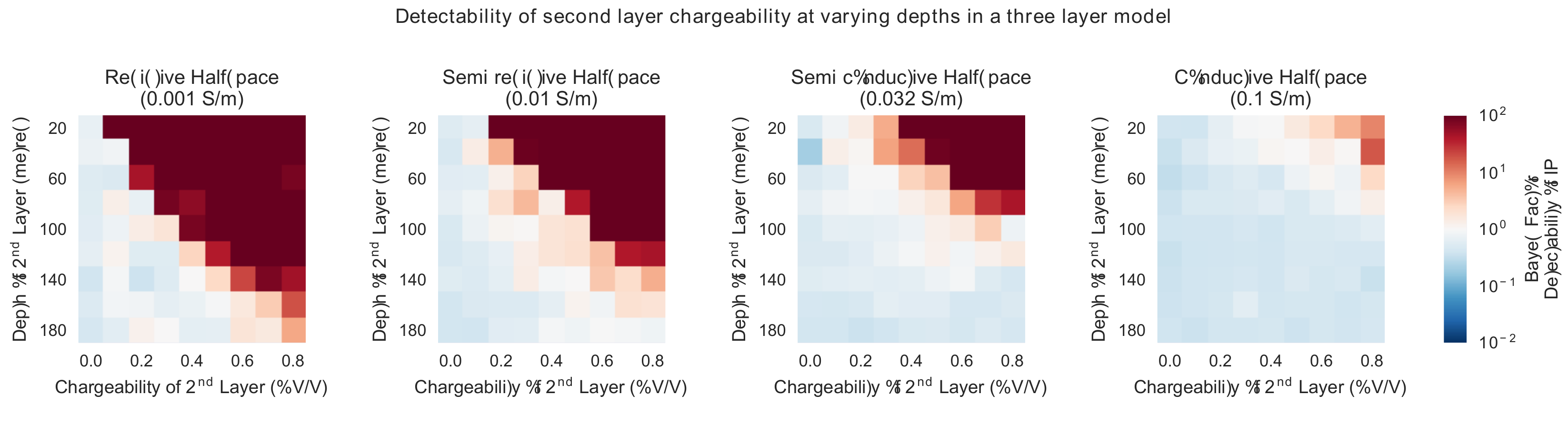}
    \caption{Shown are four case studies employing a three-layered conductivity model, each with varying chargeability and depth of the second layer, and the other layers having zero chargeability. First and second layer conductivities are 0.01 S/m and 0.1 S/m respectively, second layer thickness is 20m, and the conductivities of the the halfspace/basement layer (shown above each heatmap) reflect typical geological examples. BFIPD values above 1 indicate higher likelihood of chargeability. Within each study the heatmaps show how depth and magnitude of the chargeable layer affects the respective BFIPD statistic for each model posterior. Between studies the heatmaps show how increasing halfspace conductivity relative to the chargeable layer reduces BFIPD statistics.}
    \label{figfourstudydetectability}
\end{figure}

A comprehensive series of synthetic studies were designed to demonstrate targeted quantities that are generated by the application of RJSMC to AEM data with IP effects. The quantities of interest are detectability of chargeability, recoverability of model parameters, and goodness of fit.

For investigation of IP detectability using the BFIPD statistic from Section \ref{bfdetection}, a three-layered model with chargeability in only the middle layer was selected as basis of four almost identical case studies where only the basement conductivity varied across each of the four studies. The basement conductivities were chosen to simulate four common scenarios: a strongly resistive igneous/metamorphic basement (0.001 S/m), a hard sedimentary basement (0.01 S/m), a moderately-conductive sedimentary basement (0.032 S/m), and a conductive basement (0.1 S/m). The conductivity of the upper two layers were held constant in all cases, being  0.01 S/m and 0.1 S/m for the first and second layer respectively. The time constant and frequency dependence parameters were set to $\theta_\tau=4.07\times10^{-4}$ and $\theta_c=1.0$ respectively. The data was generated using the same VTEM ET AEM system configuration as used in \citeA{zamudio_airborne_2021} and the data noise was simulated using the noise model discussed in Section \ref{sectionlikelihood}.

Within each case study, both the depth of the upper interface and the chargeability of the second layer were jointly varied such that the BFIPD statistic could be examined as a function of the interaction of these two parameters. The BFIPD statistics were visualised as a heatmap in Figure \ref{figfourstudydetectability}. As would be expected the BFIPD statistic decreases with depth of the second layer and increases with the magnitude of second layer chargeability. Also, it can be seen that across the four studies, variation in the basement conductivity significantly affects the BFIPD statistic, where strongly resistive basement materials admit high BFIPD statistics, ranging down to low BFIPD statistics for conductive sedimentary basement materials such as shales, clays, or aquifers.

\begin{figure}
\centering
\includegraphics[width=0.95\textwidth]{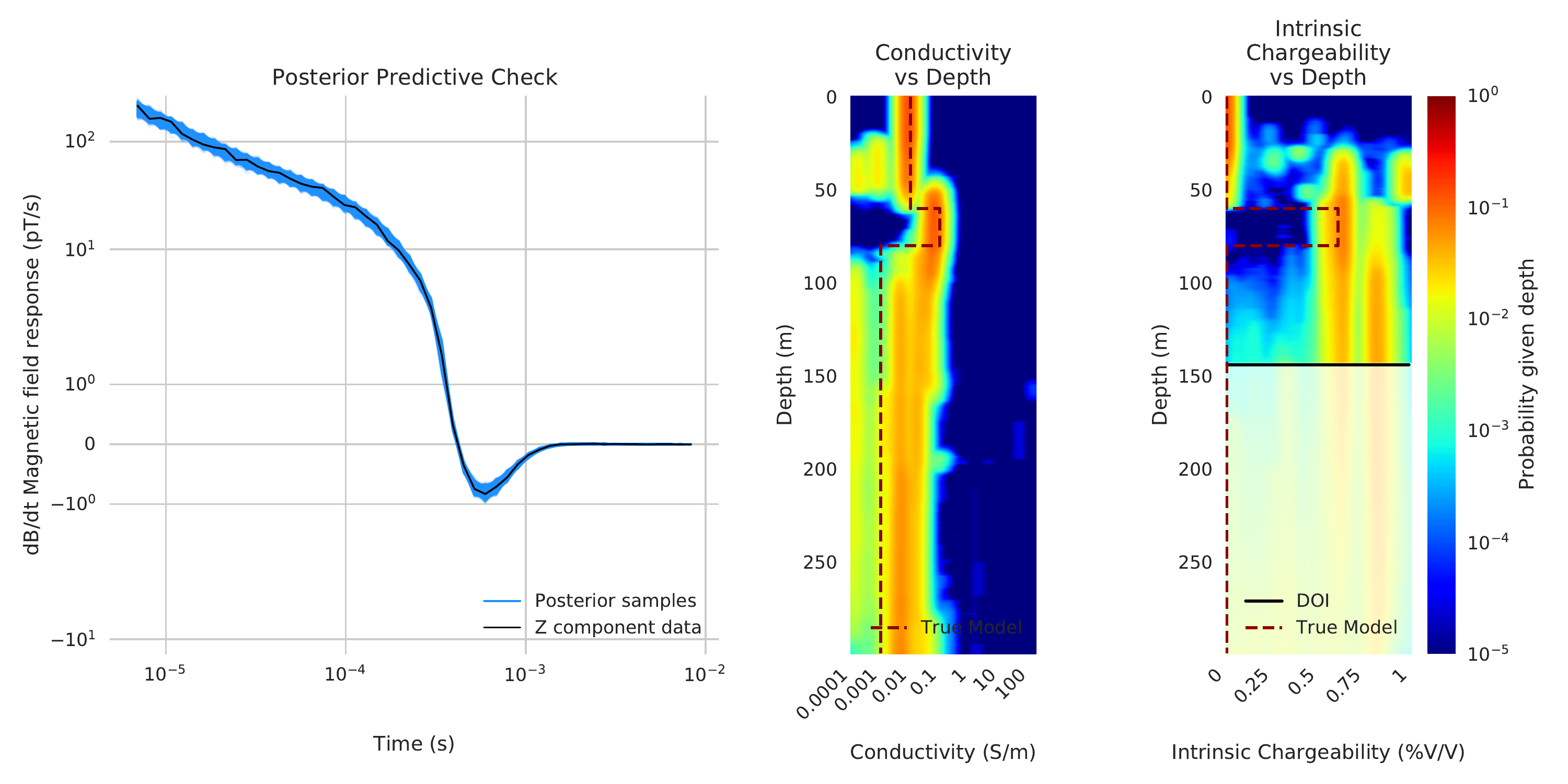}
\includegraphics[width=0.95\textwidth]{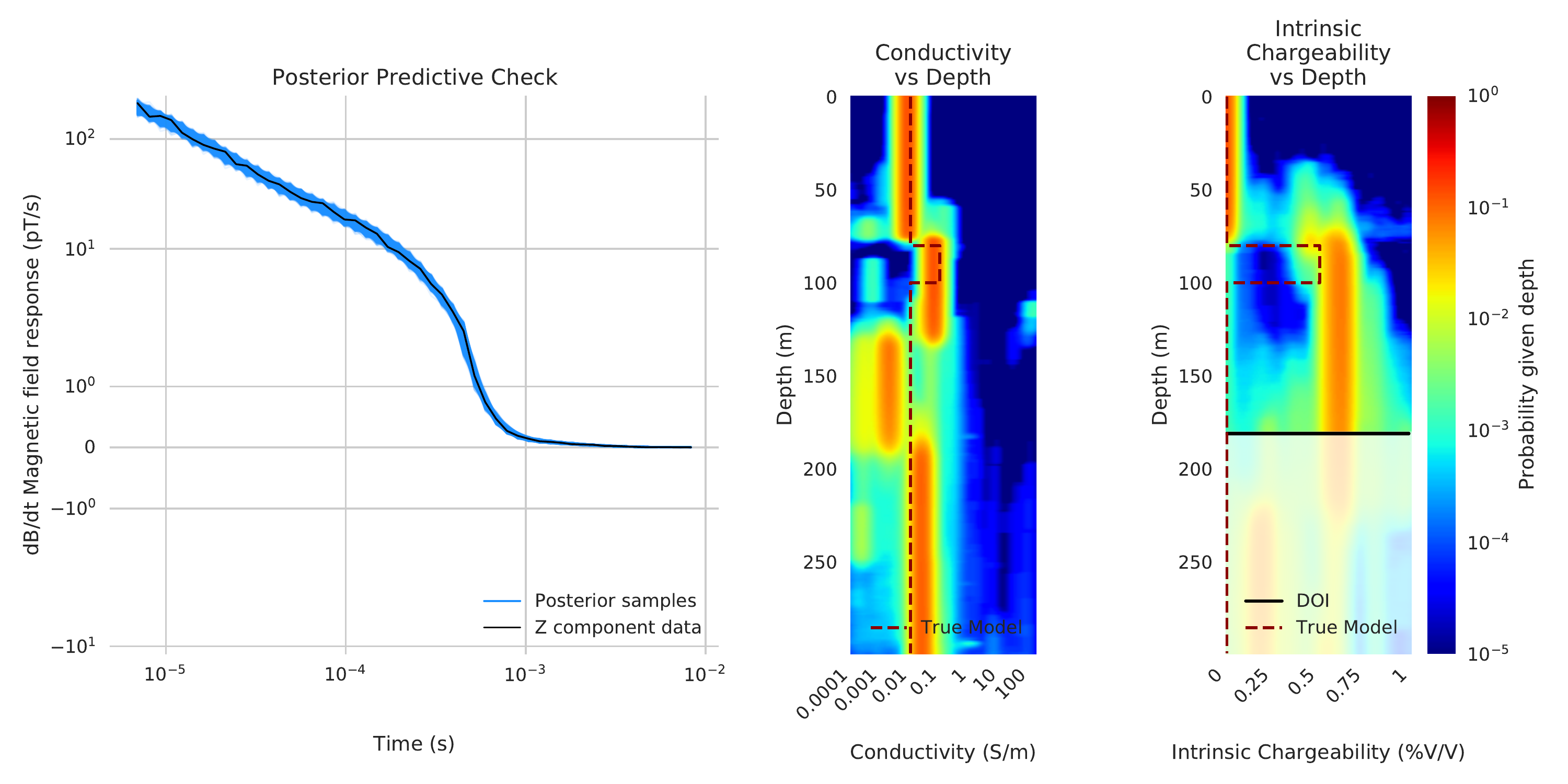}
\includegraphics[width=0.95\textwidth]{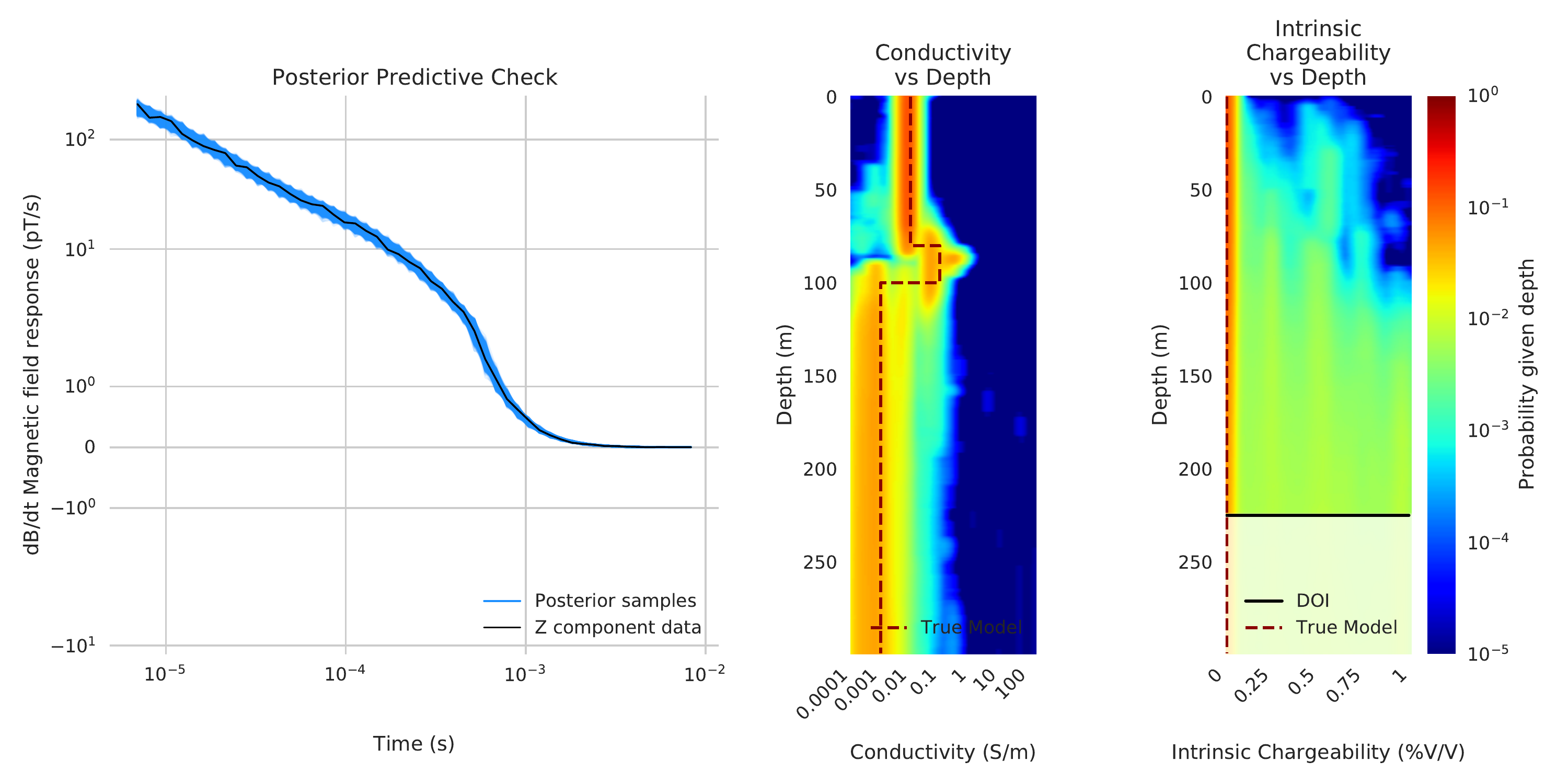}
\caption{Three figures showing data selected from the detectability synthetic study (Figure \ref{figfourstudydetectability}) to represent posterior predictive checks and parameter posteriors from data generating models with various intrinsic chargeability and depth parameters. \emph{Left:} Posterior predictive check plots where posterior samples are drawn from particles in the final target density of each inversion. \emph{Right:} Joint posterior density of the inversion of synthetic data. The true (data-generating) model parameters are identified by the dark-red dashed line in each plot. In the third figure showing the posterior of a conductive-only model the recoverability of conductivity parameters at greater depths is visibly more successful than when chargeability is present in the data-generating model.}
\label{figsynthetic}
\end{figure}

The recoverability of model parameters is well-known to be confounded by a phenomenon geophysicists term ``non-uniqueness" \cite{ellis_inversion_1998}. In the language of Bayesian statistics, this translates to the situation where the {\em maximum a posteriori} (MAP) model is not necessarily close to the {\em data-generating} model. This is frequently seen in conductivity-only inversion of AEM data and thus it is expected to be present in Bayesian AIP inference. For this investigation, we compared a selection of the model-averaged posterior densities from the four case studies outlined above against the data-generating parameters. Figure \ref{figsynthetic} shows three selected data sets from the above detectability study. The posterior is summarised in two model-averaged plots showing depth versus conductivity and depth versus intrinsic chargeability, and the ``true" data-generating model is shown on each posterior plot with a dark-red dashed line. It can be seen that the shallow layers are generally well-recovered, however the deeper layers and layer interfaces of the data-generating model are not usually reflected in the high-probability regions of the posterior. This phenomenon is well-understood for conductive-only models \cite{ellis_inversion_1998} but the extent to which it is present in models with chargeability is not well-quantified and should be the subject of further research. 

A canonical Bayesian form of goodness of fit uses the {\em posterior predictive distribution} (PPD) \cite{gelman_bayesian_2013}, which has the mathematical form
\begin{align*}
    p(\bm{y}|\Tilde{\bm{y}}) &= \int p(\Tilde{\bm{y}}|\bm{\theta})\pi(\bm{y}|\bm{\theta}) d\bm{\theta}.
\end{align*}
A visual inspection of the PPD constitutes a {\em posterior predictive check} (PPC). PPCs for posteriors generated from three synthetic data sets are shown in Figure \ref{figsynthetic}. This approach is straightforward for individual soundings; however, it can be cumbersome when examining PPDs of the thousands of soundings in a single AEM line. In such cases, it is more feasible to display summary statistics of the PPD, such as the sample mean and variance, and this approach can be seen for the Colorado study in Figure \ref{figconcharprofile}.

\subsubsection{Depth of Investigation}
A Bayesian approach to computing the level of information of the data with respect to depth, known as the depth of investigation (DOI), was developed by \citeA{blatter_trans-dimensional_2018} using entropy to measure information gain. By evaluating the continuous conductivity or chargeability parameters at discrete intervals with respect to depth, the depth of investigation divergences are computed at 1-metre intervals. It is necessary to choose a suitable divergence threshold to mask the posterior in regions that are considered to be occluded; for this work a threshold of 1 was chosen to work well in most scenarios. Figure \ref{figsynthetic} shows the DOI for the intrinsic chargeability vs depth averaged model posterior in each case. The DOI for conductivity was below the a priori maximum interface depth, indicating that in these cases the resolution of conductivity was greater than that of chargeability.

\subsection{Colorado Study}
\begin{figure}
\centering
\begin{tikzpicture}
\begin{scope}
    \node[anchor=south west,inner sep=0] (image) at (0,0) {\includegraphics[width=\textwidth]{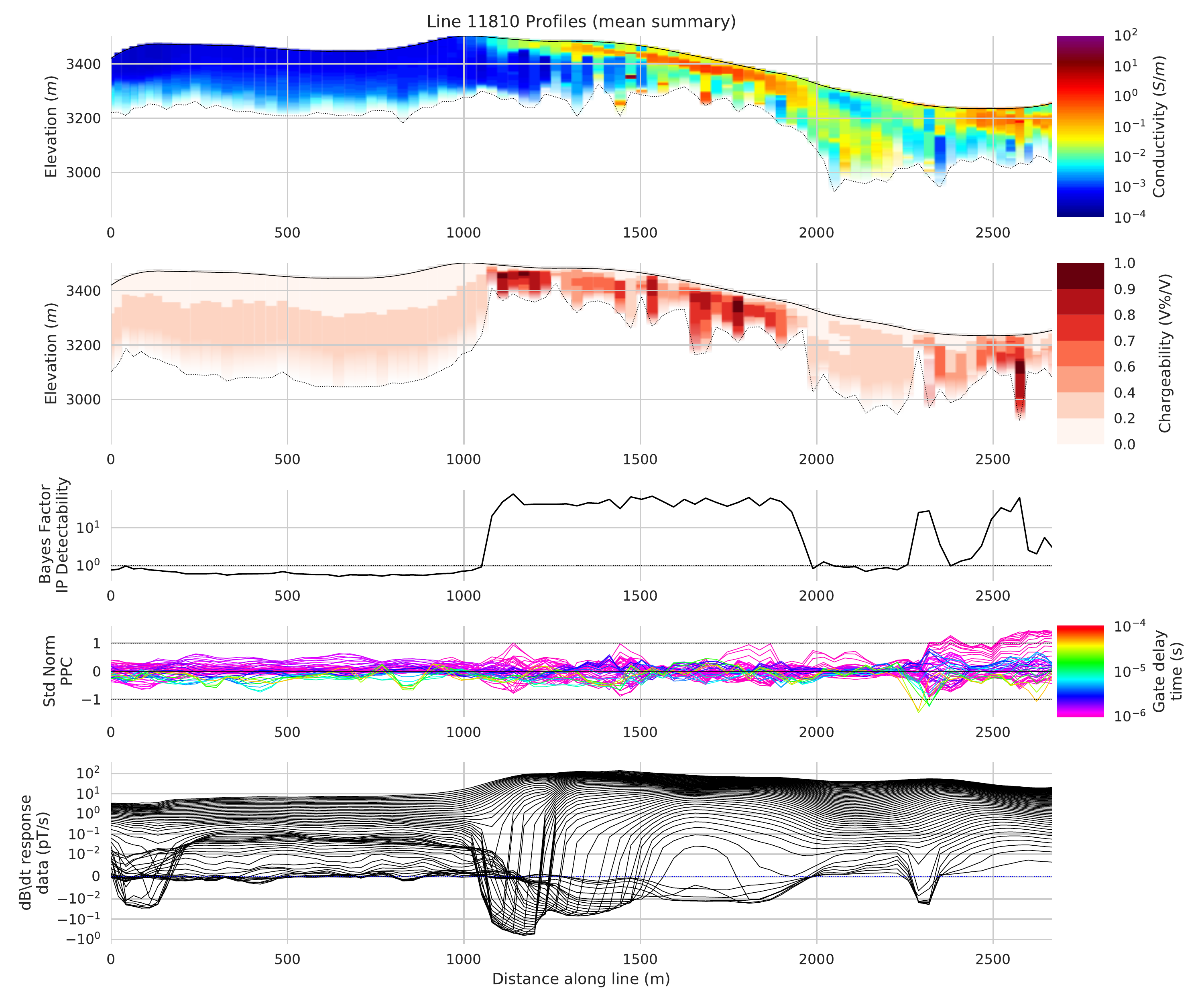}};
    \begin{scope}[x={(image.south east)},y={(image.north west)}]
        \draw[red,rounded corners=0.5] (0.430,0.05) rectangle (0.441,0.97);
    \end{scope}
\end{scope}
\end{tikzpicture}
\includegraphics[width=\textwidth]{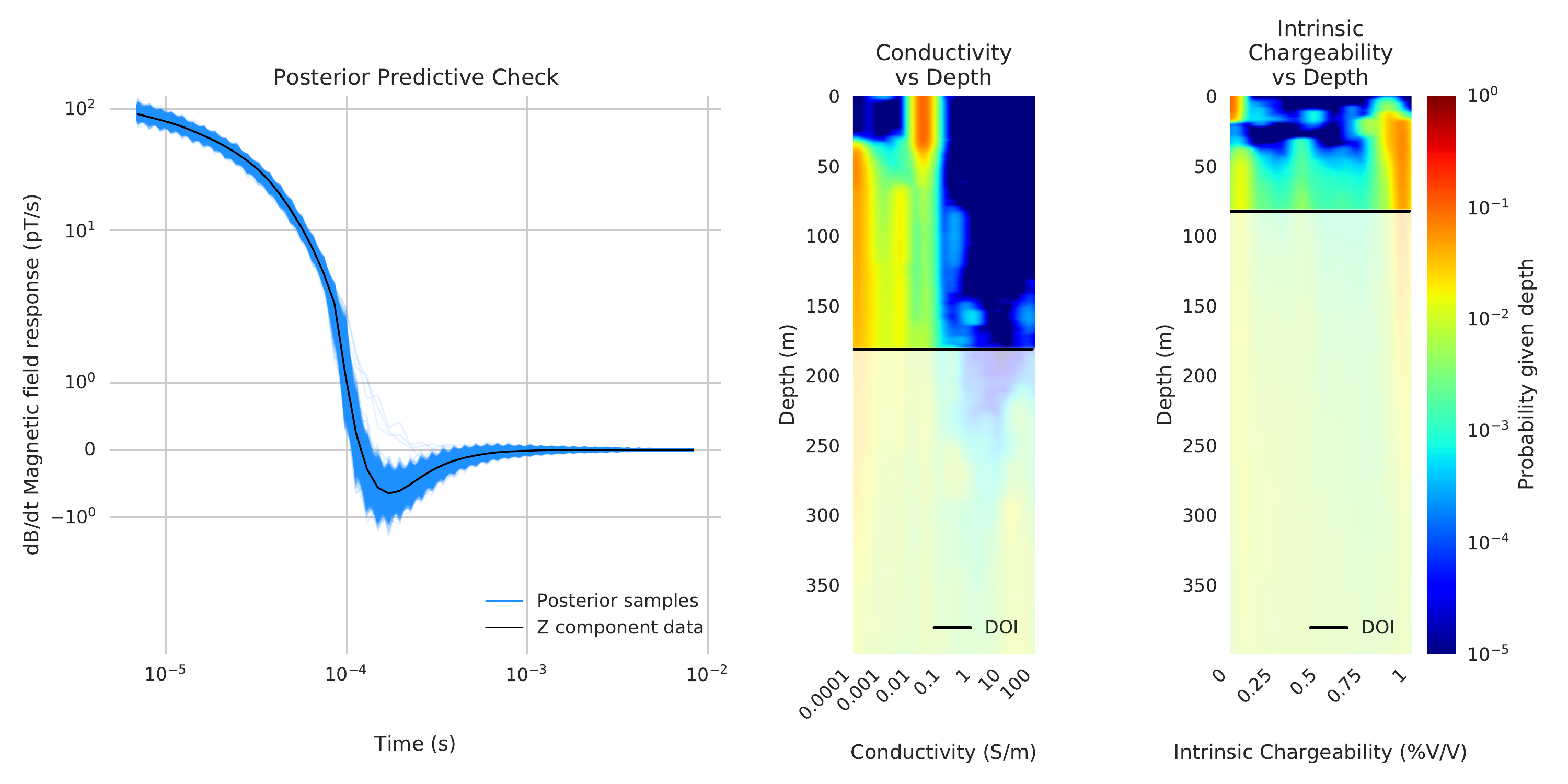}
\caption{ \emph{Top}:  Mean posterior summaries of each sounding for conductivity and chargeability parameter posteriors given VTEM ET data from the Colorado USA AEM survey. Each mean posteior summary is computed using the mean of the model-averaged parameter posterior versus depth. The third panel shows the Bayes Factor Induced Polarisation Detectability (BFIPD) statistic, which is the expected non-chargeable model probability versus the expected chargeable model probability. The fourth panel displays a summary of each data set standardised to the respective posterior predictive distribution (PPD) such that any significant deviations of the data at each gate time with respect to the PPD can be easily identified. Values between -1 and 1 fall within the standard deviations of the PPD. A selected sounding is highlighted with a red rectangle for further inspection in the bottom figure. \emph{Bottom}: Posterior predictive check and posterior vs depth plot of conductivity and chargeability for data highlighted in the red rectangle in the top figure.}
\label{figconcharprofile}
\end{figure}

The above methods were applied to parameter and model inference on one flight-line of AEM survey data from Colorado, USA (\citeA{minsley_airborne_2018}, \citeA{zamudio_airborne_2021}). Notably, there was a relatively low particle count (9600 particles per sounding) required to sufficiently permit model and parameter inference on the resultant posteriors. The selected data visibly contained negative magnetic flux density measurements that cannot be modelled using non-chargeable physics. Data was spatially decimated so that soundings are spaced by approximately $30$ metres and sections of the mean summaries are shown in Figure \ref{figconcharprofile}. Although there is no published ground truth survey for this region, it is expected that the continuous section of non-zero intrinsic chargeability in flight line 11810 is due to the frequent occurrence of disseminated pyrite present in the Mancos shale formations consistent with the surrounding geology \cite{vanderwilt_geology_1937}.

This flight line of data was also the subject of an earlier study by \citeA{viezzoli_insight_2019}. In their study, the authors compare Cole-Cole IP ground parameters, recovered via a Tikhanov regression inversion, to a geology map of the region to demonstrate spatial consistency with known geological structures in the area. Notably, their findings for MAP parameters were consistent with the results in this work, however their study did not take into account parameter and model uncertainty and did not attempt to discriminate data on the basis of the presence of induced polarisation effects.

\section{Conclusions}
In this study we demonstrated the effectiveness of Bayesian parameter inference and model inference, the latter specifically for inference on the likelihood of non-zero intrinsic chargeability when the number of layers in conductivity and chargeablility depth profiles are unknown a-priori. We have developed an SMC algorithm for inference of parameters and model probabilities that can exploit practically any parallel computing architecture, independent of the configuration of the model space and the number of particles, and is robust in the presence of pathological posteriors. In synthetic studies we have shown how well chargeability can be detected via the BFIPD statistic for a given AEM system configuration. In real data examples we have demonstrated how the BFIPD statistic is spatially consistent in a line of AEM data.

Future research stemming from this work could take many avenues. Geophysics practitioners may delve into the parameterisation and/or the assumptions of the existing noise model that was discussed in Section \ref{sectionlikelihood}. It would also be of interest to investigate the computational efficiency of cross-dimensional proposals considering that such proposal design can be enhanced with the availability of particles approximating $\pi_t$. There are potentially other efficiencies that could be implemented in the SMC algorithm itself which could quickly make it computationally competitive with well-established Bayesian inference software for geophysics. The Python 3 code that was developed for this research is part of a larger framework under development and will be released in the near future, accessible via \url{http://www.terrascope.com.au/} and as a repository at \url{https://github.com/daviesl/}.

\appendix
\section{Static Sequential Monte Carlo Algorithm}\label{staticsmcalgorithm}

\begin{algorithm}[H]
\caption{Static Sequential Monte Carlo with an MCMC Kernel}
\label{ssmcalgo}
\SetAlgoLined
\DontPrintSemicolon
    \KwOut{$\{\bm{\theta}_{T}^{(i)}\}_{i=1}^{N}$ and $\widehat{\mathcal{Z}}_{T}$}
    Initialise particles from the prior for $i=1,\dots,N$, $\bm{\theta}_{0}^{(i)}\sim\pi_0(\cdot)=p(\cdot)$.\\
    Initialise the particle weights for $i=1,\dots,N$, $W_{0}^{(i)}\leftarrow N^{-1}$.\\
    Set $t\leftarrow 0$, $\gamma_0\leftarrow 0$, and $\widehat{\mathcal{Z}}_{0}=1$.\\
    \For{$t \leftarrow 1,..,T$} {
        \textbf{\textit{Re-weight}} particles with normalised weights for $i=1,\dots,N$.
        \begin{align*}
            w_{t}^{(i)}&\leftarrow W_{t-1}^{(i)}\mathcal{L}(\bm{y}|\bm{\theta}_{t}^{(i)})^{\gamma_t-\gamma_{t-1}}\\W_{t}^{(i)}&\leftarrow \frac{w_{t}^{(i)}}{\sum_{i=1}^{N_{t}} w_{t}^{i)}}\\\widehat{\mathcal{Z}}_{t}&\leftarrow \widehat{\mathcal{Z}}_{t-1}\sum_{i=1}^{N_{t}}w_{t}^{(i)}\end{align*}\;
        \textbf{\textit{Re-sample}} $N_{t}$ particles $\bm{\theta}_{t}^{(i)*}\sim\widehat{\pi}_{t}(\cdot)$ where $\widehat{\pi}_{t}(\cdot)$ is represented by the weighted particles $\{W^{(i)}_{t},\bm{\theta}_{t}^{(i)}\}_{i=1}^{N_{t}}$\;
        \textbf{\textit{Mutate}} particles using an MCMC kernel with proposal $q(\cdot)$ \begin{align*}
        \bm{\theta}_{t+1}^{(i)}&\sim q(\cdot|\bm{\theta}_{t}^{(i)*}).
        \end{align*}\;
    }
\end{algorithm}

\section{Adapting the number of MCMC moves for multiple proposals}\label{multipleproposalR}
If we require at least one accepted proposal with a high probability greater than a tuning parameter $c'=1-c$, say $c'=0.99$, 

\begin{align}\label{requirebinaccept}
 P(X\geq 1) &\geq 1 - c
 \end{align}
 
Evaluate \ref{requirebinaccept} by taking the complement 

\begin{flalign*}
P(X\geq 1) = 1-P(X = 0) &\geq 1-c\\
P(X = 0) &\leq c\\
\text{Taking the Binomial expansion at $X=0$ }\\
(1-p_{acc})^R &\leq c\\
R\log{1-p_{acc}}&\leq \log{c}\\
R&\geq \frac{\log{c}}{\log{1-p_{acc}}}\\
\end{flalign*}

Take $R$ as the least integer upper bound

\begin{flalign}\label{drovandir}
R &= \left \lceil{\frac{\log{c}}{\log{1-p_{acc}}}}\right \rceil
\end{flalign}
For more than one proposal type and many parameters, the note the number of accepted proposals $\bm{X} \sim \mathrm{Multinomial}(R,\bm{a})$ where $\bm{X} = \{X_1,...,X_n\}$ and $\bm{a}=\{a_1,...,a_x\}$ for all parameters $i \in \{1,...,n\}$. Evaluating using the same tuning parameter $1-c$ as above,
\begin{flalign}\label{requiremultaccept}
 P(\bm{X}\geq 1) &\geq 1 - c
 \end{flalign}
Evaluate \ref{requiremultaccept} by taking the complement 
\begin{flalign*}
 P(\bm{X}\geq 1) = 1-P\left(\bigcup_{i=1}^{n}X_i = 0\right) &\geq 1-c\\
P\left(\bigcup_{i=1}^{n}X_i = 0\right) &\leq c\\
\end{flalign*}
Instead of evaluating the union of sets, it is sufficient to note that the union $\bigcup_{i=1}^{n}(X_i=0)$ bounded below by any of the events $\{X_i = 0\}$ and that the marginal distribution of the multinomial is a binomial distribution. Hence enforcing
\begin{flalign}\label{binomialconstraint}
\max\left\{P(X_i = 0)\right\}=(1-p_{acc}^{min})^R &\leq P\left(\bigcup_{i=1}^{n}X_i = 0\right) \leq c
\end{flalign}
will satisfy \ref{requiremultaccept} and as such \ref{binomialconstraint} results in the same form as \ref{drovandir} using $p_{acc}^{min}$, that is
\begin{flalign}\label{laurencer}
R &= \left \lceil{\frac{\log{c}}{\log{1-p_{acc}^{min}}}}\right \rceil
\end{flalign}
\section{Total Effective Sample Size Estimator}\label{tessappendix}
First we define the Normalised Effective Sample Size (NESS) as a value between $0$ and $1$ giving the proportional quantity of representative samples of a distribution $\pi_t(\bm{\theta}|\bm{y},k)$ that can be approximated with
\begin{align*}
    \widehat{\mathrm{NESS}}_{t,k}&\vcentcolon=\frac{\widehat{\mathrm{ESS}}_{t,k}}{N_{t,k}}.
\end{align*}
We then define the Total Effective Sample Size (TESS) as a value between $0$ and $N$ in terms of the expected NESS of all models conditional on the marginal probability of the models, namely
\begin{align*}
    \mathbb{E}[\mathrm{TESS}_t] &\vcentcolon= N\cdot\mathbb{E}[\mathrm{NESS}_{t,k}]\\
    &=N\cdot\mathbb{E}\big[\mathbb{E}[\mathrm{NESS}_{t,k}|k]\big]\tag*{(Law of Total Expectation)}\\ 
    &= N\sum_{k \in \mathcal{M}}\mathbb{E}[\mathrm{NESS}_{t,k}|k] \pi(k|\bm{y}) 
\end{align*}
Using the estimator $\widehat{\mathrm{NESS}}_{t,k}$ from above, and noting that  $\widehat{\pi}(k|\bm{y})=\frac{N_{t,k}}{N}$, we have the estimate
\begin{align*}
    \widehat{\mathrm{TESS}}_t &= N\sum_{k \in \mathcal{M}}\frac{\widehat{\mathrm{ESS}}_{t,k}}{N_{t,k}} \frac{N_{t,k}}{N}= \sum_{k \in \mathcal{M}}\widehat{\mathrm{ESS}}_{t,k}
\end{align*}

\section{Ratio of Normalising Constants Between Successive Target Densities}\label{adaptiverjmcmcappendix}
This appendix demonstrates that the ratio of normalising constants between successive target densities in RJSMC is derived in a similar way to those for single model static SMC. Here we will use the notation $p_t(\bm{\theta}|k)$ to represent the conditional (normalised) target distribution. Noting $\eta_{t,k}(\bm{\theta}_k|k)=Z_{t,k}p_t(\bm{\theta}|k)$, the ratio of normalising constants can be found as follows
\begin{align*}
    \mathcal{Z}_{t,k}
        &=\int_{\bm{\theta}_k}\eta_{t,k}(\bm{\theta}_k|k)d\bm{\theta}\\
        &=\int_{\bm{\theta}}\frac{\eta_{t,k}(\bm{\theta}_k|k)}{p_{t-1}(\bm{\theta}_k|k)} p_{t-1}(\bm{\theta}_k|k) d\bm{\theta}\\
        &=\int_{\bm{\theta}}\mathcal{Z}_{t-1,k}\frac{\eta_{t,k}(\bm{\theta}_k|k)}{\eta_{t-1,k}(\bm{\theta}_k|k)} p_{t-1}(\bm{\theta}_k|k) d\bm{\theta}\\
        \frac{\mathcal{Z}_{t,k}}{\mathcal{Z}_{t-1,k}}&=\int_{\bm{\theta}}\frac{\eta_{t,k}(\bm{\theta}_k|k)}{\eta_{t-1,k}(\bm{\theta}_k|k)} p_{t-1}(\bm{\theta}_k|k) d\bm{\theta}.
\end{align*}
\noindent Now using the weighted sample $\{W_{t-1}^{(i,k)},\bm{\theta}_{t-1}^{(i,k})\}_{i=1}^{N_{t,k}}$ from $p_{t-1}(\bm{\theta}_k|k)$ we obtain the following Monte Carlo estimate
\begin{align*}
    \widehat{\frac{\mathcal{Z}_{t,k}}{\mathcal{Z}_{t-1,k}}}&=\sum_{i=1}^{N_{t,k}}W_{t-1}^{(i,k)}\frac{\eta_{t,k}(\bm{\theta}^{(i)}|k)}{\eta_{t-1,k}(\bm{\theta}^{(i)}|k)}.
\end{align*}
In the case where we are using a RJMCMC kernel, the term $\eta_{t,k}(\bm{\theta}^{(i)}|k)/\eta_{t-1,k}(\bm{\theta}^{(i)}|k)$ is the incremental weight. In this instance, the estimate of the ratio of normalising constants is 
\begin{align*}
    \widehat{\frac{\mathcal{Z}_{t,k}}{\mathcal{Z}_{t-1,k}}}=\sum_{i=1}^{N_{t,k}} w_{t,k}^{(i)}.
\end{align*}
\section{Reversible jump MCMC proposals}
Using $\pi(\bm{\theta}_k,k|\bm{y})$ as the target of the state space $\bm{\theta}=\bigcup_{k\in\mathcal{K}}(\{k\}\times\mathcal{R}^{N_{t,k}})$, RJMCMC proposal construction requires the following design choices:
  \begin{enumerate}
        \item \textbf{Dimension Match}: Given $n_k=|M_k|$, $n_{k'}=|M_{k'}|$, draw random variables $\bm{u_k}\sim g_k(\cdot)$, $\bm{u_{k'}}\sim g_{k'}(\cdot)$ of length $w_k$ and $w_{k'}$ such that
        \begin{align*}
            n_k + w_k &= n_{k'} + w_{k'}.
        \end{align*}
        \item \textbf{Bijective Map}: $h_{k\rightarrow k'}:\mathcal{R}^{n_k}\times\mathcal{R}^{w_k}\rightarrow\mathcal{R}^{n_{k'}}\times\mathcal{R}^{w_{k'}}$ is chosen to map
        \begin{align*}
            \bm{\theta}_{k'}'&= h_{k\rightarrow k'}(\bm{\theta}_k,\bm{u})
        \end{align*}
  \end{enumerate}
  
\noindent The acceptance ratio that satisfies detailed balance is
\begin{align*}
    \alpha [(k,\bm{\theta}_k),(k',\bm{\theta}_{k'}')]&=1\wedge\frac{\pi(k',\bm{\theta}_{k'}'|\bm{y})q_{k'\rightarrow k}(\bm{\theta}_{k},k|\bm{\theta}_{k'}',k')g_{k'\rightarrow k}(\bm{u}_{k'})}{\pi(k,\bm{\theta}_{k}|\bm{y})q_{k\rightarrow k'}(\bm{\theta}_{k'}',k'|\bm{\theta}_{k},k)g_{k\rightarrow k'}(\bm{u}_k)}\Bigg|\frac{\partial_{h_{k\rightarrow k'}}(\bm{\theta}_k,\bm{u})}{\partial(\bm{\theta}_k,\bm{u})}\Bigg|.
\end{align*}

\subsection{RJMCMC Proposals for the 1D layered Earth model}\label{rjmcmclayereddesign}
\cite{malinverno_parsimonious_2002} introduced a proposal for the 1D layered model based on the change-point model from \cite{green_reversible_1995}. This model essentially specifies the model space as an arbitrary number of identically-distributed change-points representing layer interfaces, each with associated properties. The original definition used the so-called ``grid trick" to enable birth-death proposals of layer interfaces. This was formalised instead by \cite{dosso_efficient_2014} to use a Dirichlet-type prior on the layers themselves, visualised as the homogeneous space between the layer interfaces. The priors are
\begin{align*}
    \text{Number of layers : } P(k) &= \frac{1}{k_\mathrm{max}-k_\mathrm{min}}\\
    \text{i}^{th}\text{-layer associated physical property j : } P(\beta_{i,j}|k) &=  \frac{1}{\beta_{\mathrm{max}}-\beta_{\mathrm{min}}}\\
    \text{Layer thickness : } P(\bm{z}_k|k) &= \frac{k!}{z_{\mathrm{max}}^{k}}\\
    \text{i}^{th}\text{-layer parameter }\bm{\theta}_{i,k} &= \{\beta_{i,0},\dots,\beta_{i,J},z_i\}
\end{align*}
The combined prior is given by
\begin{align*}
    P(k)P(\bm{\theta}_k|k)&=\frac{1}{ k_\mathrm{max}-k_\mathrm{min}}\frac{P(\bm{z}_k|k)}{\prod^J_{i=1} (\beta_{\mathrm{max}}-\beta_{\mathrm{min}})^{k+1}}
\end{align*}
The birth acceptance term is comprised of the prior and proposal ratios. The prior ratio is
\begin{align*}
    \frac{P(k')P(\bm{\theta}_{k'}|k')}{P(k)P(\bm{\theta}_k|k)}&=\frac{k+1}{z_{\mathrm{max}}\prod^J_{i=1} (\beta_{\mathrm{max}}-\beta_{\mathrm{min}})}
\end{align*}
The proposal ratio for inserting a new layer in the $i^{th}$ position (or a uniform probability of inserting a layer in the interval $(0,z_{\mathrm{max}})$) combined with the reverse step (death of the proposed layer, with probability $1/(k+1)$) is
\begin{align*}
    \frac{Q(k,\bm{\theta}_k|k',\bm{\theta}_{k'})}{Q(k',\bm{\theta}_{k'}|k,\bm{\theta}_k)}&=\frac{1}{k+1}\frac{z_{\mathrm{max}}}{Q_\beta( \bm{\beta}'|\bm{\beta})}
\end{align*}
The acceptance term for layer birth, combining all terms, is
\begin{align*}
    A_{\mathrm{birth}}[(k,\bm{\theta}_k),(k',\bm{\theta}_{k'})]&=1\wedge\frac{1}{Q_\beta( \bm{\beta}'|\bm{\beta})}\frac{1}{\prod^J_{i=1} (\beta_{\mathrm{max}}-\beta_{\mathrm{min}})}\frac{\mathcal{L}(\bm{y}|k',\bm{\theta}_{k'}')}{\mathcal{L}(\bm{y}|k,\bm{\theta}_{k})}.
\end{align*}
If $Q_\beta( \bm{\beta}'|\bm{\beta})$ is chosen to be simply drawing from the uniform prior for $\beta_i$ for the $i^{th}$ layer being inserted, the acceptance ratio reduces to the likelihood ratio
\begin{align*}
A_\text{Na\"ive} [(k,\bm{\theta}_k),(k',\bm{\theta}_{k'}')]
&=
1\wedge\frac{\mathcal{L}(\bm{y}|k',\bm{\theta}_{k'}')}{\mathcal{L}(\bm{y}|k,\bm{\theta}_{k})}.
\end{align*}

\subsection{An adaptive reversible jump MCMC proposal for the 1D layered Earth model}
A common ailment of RJMCMC proposals is the poor mixing phenomenon identified by very low acceptance rates. The design of adaptive proposals is a vast field of research and is naturally problem specific. In this appendix we will identify one possible construction for an adaptive RJMCMC proposal which takes advantage of the availability of $\pi_t$ in an SMC algorithm without guarantees for performance in any particular context.

This proposal design follows that of the birth/death design in Appendix \ref{rjmcmclayereddesign} where the vector of auxilliary variables in the birth move is $\bm{u}=[u_d,u_p]$, and the depth of the new layer $u_d$ is first drawn independently and uniformly over the range of the allowed depths. Note that since layer interfaces are sorted in order of depth, we can find the index $i$ of the layer interface being inserted. 
Following this, we consider that the second auxilliary variable is drawn as from a standard Gaussian and then transformed via a bijective map. Thus we consider below the construction of $h_{k\rightarrow k'}(\bm{\theta}_k,\bm{u})$.

If it can be assumed that $\pi_t(\bm{\theta}_t|k)$ is unimodal and approximately Gaussian (an assumption that rarely holds), its variance can be approximated via the sample variance
\begin{flalign*}
\bm{\Sigma}_{k,t}&=\frac{1}{N_{t,k}-1}\sum_{i=1}^{N_{t,k}}(\bm{\theta}_{k,t}^{(i)}-\mu_{k,t})(\bm{\theta}_{k,t}^{(i)}-\mu_{k,t})^{T}
\end{flalign*}
We choose the bijective transform for the birth of one layer to be
\begin{align*}
    \bm{\theta}_{k+1,t}[j]&=\bm{\theta}_{k,t}[h(j)]+\bm{\Sigma}_{k+1,t}^{\frac{1}{2}}[i,j]u
\end{align*}
\noindent where $j$ indexes each parameter component independently, $h(j)$ maps the indices of $\bm{\theta}_k$ to $\bm{\theta}_{k+1}$, and $\bm{\Sigma}_{k+1,t}^{\frac{1}{2}}$ is the ZCA colouring matrix (square root) of $\bm{\Sigma}_{k+1,t}$. Also note that the row index $i$ is essentially arbitrary, however a common choice is the index of the layer being inserted into the model.

To form the Jacobian, we take the partial derivatives
\begin{align*}
    \frac{\partial\bm{\theta}_{k+1}[j]}{\partial \bm{\theta}_{k}[h(j)]}&=1\\
    \frac{\partial\bm{\theta}_{k+1}[j]}{\partial u}&=\bm{\Sigma}_{k+1,t}^{\frac{1}{2}}[i,j]\\
\end{align*}
The Jacobian becomes a mostly triangular matrix with one row replicating the non-zero element locations. It can be re-arranged to a block triangular matrix
\begin{align*}
    \bm{J}&=\begin{bmatrix}
    \bm{A} & \bm{B}\\
    0 & \bm{D}
    \end{bmatrix}
\end{align*}
where $\bm{A}$ is diagonal and comprised of $\frac{\partial\bm{\theta}_{k+1}[j]}{\partial \bm{\theta}_{k}[h(j)]}=1$ entries, and
\begin{align*}
     \bm{D}&=\begin{bmatrix}
    1 & \frac{\partial\bm{\theta}_{k+1}[h^{-1}(j)]}{\partial u_p}\\
    1 & \frac{\partial\bm{\theta}_{k+1}[j]}{\partial u_p}
    \end{bmatrix}.
\end{align*}
Hence the determinant becomes
\begin{align*}
    \det{\bm{J}}&=\det{\bm{A}}\det{\bm{D}}\\
    &=\frac{\partial\bm{\theta}_{k+1}[j]}{\partial u_p} - \frac{\partial\bm{\theta}_{k+1}[h^{-1}(j)]}{\partial u_p}\text{\  as }\det{\bm{A}}=1\\
    &=\bm{\Sigma}_{k+1,t}^{\frac{1}{2}}[i,j]-\bm{\Sigma}_{k+1,t}^{\frac{1}{2}}[i,h(j)].
\end{align*}
For the death move, we need to solve for $u_p$. Using the knowledge that elements $\bm{\theta}_{k}[j]$ and $\bm{\theta}_{k}[h(j)]$ are equal for $j\neq h(j)$, we derive
\begin{align*}
    u_p&=\frac{\bm{\theta}_{k+1}[j]-\bm{\theta}_{k+1}[h(j)]}{\bm{\Sigma}_{k+1,t}^{\frac{1}{2}}[i,j]-\bm{\Sigma}_{k+1,t}^{\frac{1}{2}}[i,h(j)]}.
\end{align*}
\bsp 

\acknowledgments

%
%
%
%
%
%

This work was financially supported by the Australian Government Research Training Program, the Queensland University of Technology Centre for Data Science, and an Australian Research Council Discovery Project (DP200102101). Computing resources were provided by Queensland University of Technology. The authors would like to give special thanks to Dr Andrea Viezzoli of Aarhaus Geophysics for crucial advice concerning the Colorado data,
and to Dr Ross Brodie of Geoscience Australia for providing the forward model code for IP in a TDEM system (see the development branch in \citeA{brodie_geoscience_2016}).


AEM Data from the Colorado case study was first presented in \cite{minsley_airborne_2018} and will be available under \citeA{zamudio_airborne_2021} as an entry in the data repository website \url{http://www.sciencebase.gov}.

\bibliography{references.bib}
\label{lastpage}

\end{document}